\documentclass[journal=jctcce,manuscript=article,email=false]{achemso}

\usepackage{amsmath,amsfonts,amssymb}
\usepackage{array, booktabs, tabularx}
\usepackage{xcolor, graphicx, tikz}
\usepackage[caption=false]{subfig}
\usepackage{hyperref} 
\usepackage{braket}
\usepackage{bm}
\usepackage[normalem]{ulem} 

\hypersetup{colorlinks = True, allcolors = {blue}}

\graphicspath{{Figures/}}

\newcommand{\BC}[1]{\hat{a}_{#1}^\dagger}
\newcommand{\BA}[1]{\hat{a}_{#1}}
\newcommand{\BN}[1]{\hat{n}_{#1}}

\newcommand{\BP}[1]{\hat{p}_{#1}}
\newcommand{\BQ}[1]{\hat{q}_{#1}}

\def\Ham{ \hat{H} }

\newcommand{\Eq}[1]{Eq.~({#1})}
\newcommand{\Fig}[1]{Figure~{#1}}

\newcommand{\Sec}[1]{Section~{#1}}
\newcommand{\Table}[1]{Table~{#1}}
\newcommand{\Reference}[1]{Ref.~{#1}}

\title{
Simulating Chemistry on Bosonic Quantum Devices 
}

\author{ \small{
Rishab Dutta,$^\dagger$
Delmar G. A. Cabral,$^\dagger$ 
Ningyi Lyu,$^\dagger$
Nam P. Vu,$^{\ddagger, \dagger}$
Yuchen Wang,$^{\P}$
Brandon Allen,$^{\dagger}$
Xiaohan Dan,$^{\dagger}$
Rodrigo G. Corti\~{n}as,$^{\S, \Vert}$
Pouya Khazaei,$^{\perp}$
Max Sch{\"a}fer,$^{\S, \Vert}$
Alejandro C. C. d. Albornoz,$^{\S, \Vert}$
Scott E. Smart,$^{\#}$
Scott Nie,$^{\#}$
Michel H. Devoret,$^{\S, \Vert}$
David A. Mazziotti,$^{@}$ 
Prineha Narang,$^{\#}$ 
Chen Wang,$^{\triangle}$ 
James D. Whitfield,$^{\nabla}$ 
Angela K. Wilson,$^{\dagger \dagger}$ 
Heidi P. Hendrickson,$^{\ddagger}$ 
Daniel A. Lidar,$^{\ddagger \ddagger}$ 
Francisco P\'erez-Bernal,$^{\P \P, \S \S}$ 
Lea F. Santos,$^{\Vert \Vert}$ 
Sabre Kais,$^{\P}$ 
Eitan Geva,$^{\perp}$ 
Victor S. Batista$^{\dagger, \Vert}$
}}

\affiliation{ \small{
$^\dagger$Department of Chemistry, Yale University, New Haven, CT 06520, USA; 
$^{\ddagger}$Department of Chemistry, Lafayette College, Easton, PA 18042, USA;
$^{\P}$Department of Chemistry, Department of Physics, and Purdue Quantum Science and Engineering Institute, Purdue University, West Lafayette, IN 47907, USA;
$^{\S}$Department of Applied Physics and Department of Physics, Yale University, New Haven, CT 06520, USA;
$^{\Vert}$Yale Quantum Institute, Yale University, New Haven, CT 06511, USA;
$^{\perp}$Department of Chemistry, University of Michigan, Ann Arbor, MI 48109, USA;
$^{\#}$Division of Physical Sciences, College of Letters and Science and Department of Electrical and Computer Engineering, University of California, Los Angeles, Los Angeles, CA 90095, USA;
$^{@}$Department of Chemistry and The James Franck Institute, The University of Chicago, Chicago, IL 60637, USA;
$^{\triangle}$Department of Physics, University of Massachusetts-Amherst, Amherst, MA 01003, USA;
$^{\nabla}$Department of Physics and Astronomy, Dartmouth College, Hanover, NH 01003, USA
$^{\dagger \dagger}$Department of Chemistry, Michigan State University, East Lansing, MI 48864, USA;
$^{\ddagger \ddagger}$Department of Electrical \& Computer Engineering, Department of Chemistry, Department of Physics \& Astronomy, and Center for Quantum Information Science \& Technology, University of Southern California, Los Angeles, CA 90089, USA;
$^{\P \P}$Departamento de Ciencias Integradas y Centro de Estudios Avanzados en F\'isica, Matem\'aticas y Computaci\'on, Universidad de Huelva, Huelva 21071, Spain;
$^{\S \S}$Instituto Carlos I de F\'isica Te\'orica y Computacional, Universidad de Granada, Granada 18071, Spain;
$^{\Vert \Vert}$Department of Physics, University of Connecticut, Storrs, Connecticut 06269, USA
}}

\AtBeginDocument{\AtBeginShipoutNext{\AtBeginShipoutDiscard}}
\addtocounter{page}{-1}

\begin{document}


\pagebreak
\begin{abstract}

Bosonic quantum devices offer a novel approach to realize quantum computations, where the quantum two-level system ({\em qubit}) is replaced with the quantum (an)harmonic oscillator ({\em qumode}) as the fundamental building block of the quantum simulator. 
The simulation of chemical structure and dynamics can then be achieved by representing or mapping the system Hamiltonians in terms of bosonic operators. 
In this perspective, we review
recent progress and future potential of using bosonic quantum devices 
for addressing a wide range of challenging chemical problems, including the calculation of molecular vibronic spectra, the simulation of gas-phase and solution-phase adiabatic and nonadiabatic chemical dynamics, the efficient solution of molecular graph theory problems, and the calculations of electronic structure. 

\end{abstract}


\section{Introduction}

Computational chemistry has made significant progress in solving a wide range of chemical problems using classical computers. 
However, many critical chemical phenomena remain beyond what classical computers can efficiently simulate. 
This state of affairs, which originally motivated Feynman to propose quantum computers, \cite{Feynman2018simulating} 
also motivated developing hardware and software for performing chemical calculations on quantum computing platforms. \cite{Lidar1999calculating}
This has led to several proposals of hybrid algorithms that divide the computational task between classical and 
quantum computing platforms in a manner that is meant to offer a quantum advantage. \cite{Cao2019,Bauer2020,McArdle2020,Motta2022}
Examples include algorithms for 
calculating ground and excited state electronic structure of molecules,
\cite{Peruzzo2014,Grimsley2019,Mcardle2019,Motta2020,Smart2021,Kyaw2023} 
for predicting properties of molecular systems in thermal equilibrium, \cite{Motta2020} and for simulating chemical dynamics. \cite{Ollitrault2021,Wang2023GQME} 
Most of those algorithms have been designed for quantum computing platforms that are based on highly controllable coupled arrays of two-level systems, or {\em qubits}. 
However, persistent limitations on the circuit depth imposed by currently available qubit-based 
noisy intermediate-scale quantum (NISQ) hardware have created a bottleneck that so far has prevented the demonstration of the desired quantum advantage. 

The recent emergence of highly controllable bosonic quantum devices opens the door for a potentially paradigm-shifting alternative 
quantum hardware platform for computational chemistry.
Unlike qubit-based quantum platforms, the fundamental building block of these bosonic devices is a harmonic oscillator, known as a {\em qumode}. 
Unlike the qubit whose state lives in a 2D Hilbert space, the state of a qumode lives in an 
infinite dimensional Hilbert space which can be 
represented in either the discrete basis of the harmonic oscillator stationary states or a continuous basis of the position or momentum operators' eigenfunctions. 
One potential advantage of qumodes over qubits is that the expanded Hilbert space dimensionality is likely to lead to a reduction in circuit depth. 
Another potential advantage is that the simulation of chemical structure and dynamics is often based on Hamiltonians that are either given in terms of or can be mapped onto bosonic operators, which makes working in the oscillator space a natural choice.    

In this paper, we focus on a promising experimental realization of a qumode-based platform that uses circuit quantum electrodynamics (cQED). \cite{Copetudo2024}
In this case, superconducting resonators serve as qumodes, 
with the interactions between resonators controlled via an auxiliary superconducting circuit. \cite{Deleglise2008,Hacker2019,Reagor2016}
Other hardware approaches to bosonic quantum computing include arrays of trapped ions, \cite{Bruzewicz2019,fluhmann2019encoding} optical modes, \cite{Huh2015} or mechanical resonators. \cite{Chu2020}

The remainder of this perspective is organized as follows.
An introduction to 
bosonic quantum computing is provided in \Sec{\ref{sec: background}}. 
Recent progress in simulating molecular vibronic spectra\cite{Wang2020,Chavez2023} and chemical quantum dynamics \cite{Lyu2023} on bosonic quantum devices is reviewed in \Sec{\ref{sec: past_work}}. 
The potential for bosonic quantum devices to solve a wide range of timely problems in computational chemistry, including nonadiabatic chemical
dynamics, the efficient solution of molecular graph theory problems and electronic
structure calculations
is highlighted in \Sec{\ref{sec: current_work}}.
Concluding remarks are provided in \Sec{\ref{sec: conclusions}}.


\section{Bosonic quantum computing} \label{sec: background}

In this section, we introduce some basic aspects of bosonic systems and then discuss how to realize bosonic quantum computing with resonators coupled to transmon qubits.

\subsection{A primer on bosonic systems}


\begin{figure}[t]
\centering
\includegraphics[width=0.9\columnwidth]{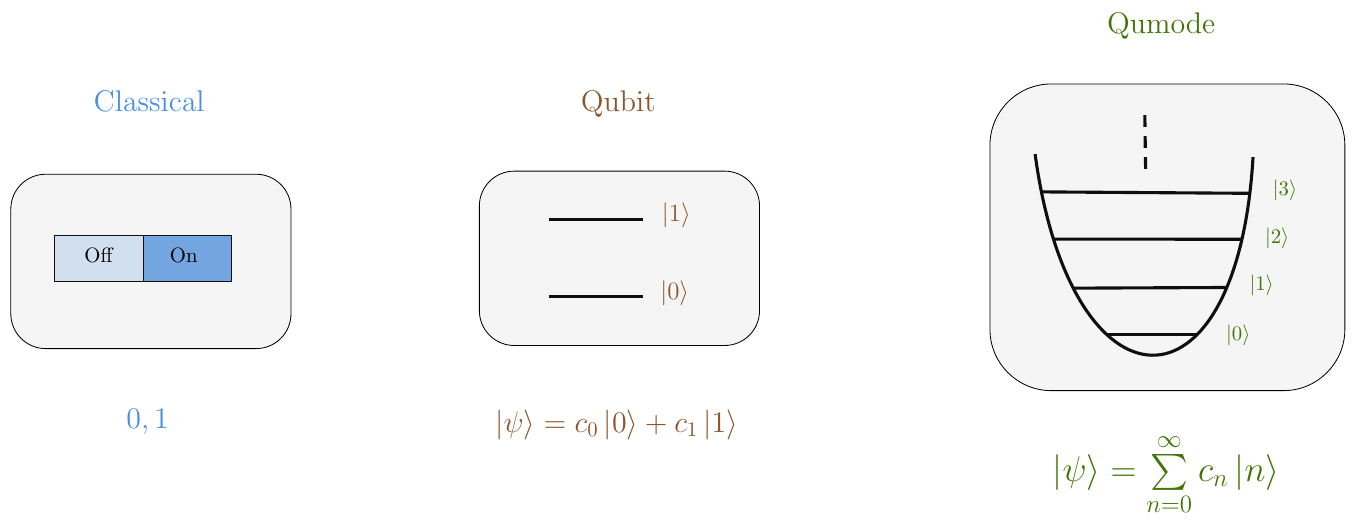}
\caption{
    Comparisons between classical bits and quantum information as represented by qubit and qumode. 
    Classical computers process information as combinations of 0 (on) and 1 (off). 
    A qubit is a quantum two-level quantum system, whose state can be given as a superposition of the two stationary states $\{ \ket{0},\ket{1} \}$. 
    A qumode is a quantum harmonic oscillator, whose state can be given by a superposition of the infinitely many stationary states, $\{ \ket{0},\ket{1},\ket{2},\ket{3}, \cdots \}$.
}
\label{fig: quantum_information}
\end{figure}


Consider a system of $N$ 
quantum harmonic oscillators
(QHOs). 
Each QHO corresponds to a bosonic mode, which we will refer to as a qumode.
The Hilbert space of a single qumode is spanned by a countably infinite Fock basis $\{ \ket{n} | \: n \in \mathbb{Z}_{\geq 0} \}$. 

The elementary bosonic creation and annihilation operators for a single qumode are defined as: 
\begin{subequations}
\begin{align}
\BC{} \ket{n} 
&\equiv \sqrt{n + 1} \ket{n + 1},
\\
\BA{} \ket{0} 
&\equiv 0, \quad 
\BA{} \ket{n} 
\equiv \sqrt{n} \ket{n - 1} \: (n > 1).
\end{align}    
\end{subequations}
Here, $\BC{}$ and $\BA{}$ are the creation and annihilation operators, respectively, and $\ket{0}$ is the ground state. 
An arbitrary Fock basis state can then be written as 
\begin{equation} \label{eq: singlemode_fock_state}
\ket{n}
= \frac{1}{\sqrt{n!}} \: ( \BC{} )^n \ket{0},
\end{equation}
which is an eigenstate of the number operator, $\hat{n} = \BC{} \BA{} $:
\begin{equation}
\BC{} \BA{} \ket{n} = \hat{n} \ket{n} = n\ket{n}.
\label{eq:number}
\end{equation}
The number operator represents an observable corresponding to the number of photons in a single mode, or equivalently the excitation level of the QHO mode.
Thus, \Eq{\ref{eq: singlemode_fock_state}} can practically represent the quantum information unit known as a \textit{qudit} when truncated at a finite level, which generalizes the concept of a qubit, where $\ket{n}$ is either $\ket{0}$ or $\ket{1}$. 
A schematic comparing a classical bit, a qubit, and a qumode is presented in \Fig{\ref{fig: quantum_information}}.

The Hilbert space of the multi-qumode system 
is given by the tensor product of the single-qumode Hilbert spaces.
Thus, the Fock basis states 
$ \ket{\mathbf{n}} \equiv \ket{n_1, n_2, \cdots, n_N} $ 
can be written as 
\begin{equation} \label{eq: multimode_fock_state}
\ket{\mathbf{n}}
= \ket{n_1} \otimes \ket{n_2} \otimes \cdots \otimes \ket{n_N},
\end{equation} 
where $\otimes$ denotes the tensor product and $\ket{n_j}$ is the stationary state of the $j$-th qumode. 
The corresponding bosonic creation and annihilation operators, $\{ \BC{j} , \BA{j} \}$ satisfy the canonical commutation relations: 
\begin{equation} \label{eq: ccr}
[ \BC{j}, \BC{k} ]  
= 0, \quad
[ \BA{j}, \BC{k} ]  
= \delta_{jk}.
\end{equation}


\begin{table}[t]
    \centering
    \begin{tabular}{cccc}
        \hline 
        Gate && Unitary operation 
        \\ \hline 
        Displacement && 
        $ D_j (\alpha) 
        = \exp{ ( \alpha \: \BC{j} - \alpha^* \: \BA{j}) } $ 
        \\
        Rotation && $R_j (\phi) 
        = \exp{ ( i \phi \: \BN{j} ) } $ 
        \\
        Squeezing && 
        $ S_j (z) 
        = \exp{ \Big[ \frac{1}{2} \: \big( z^* \: \BA{j}^2 -z \: ( \BC{j} )^2 \big) \Big] }$ 
        \\
        Beamsplitter && 
        $ BS_{jk} ( \theta, \phi ) 
        = \exp{ \Big[ \theta \: ( e^{i\phi} \: \BC{j} \BA{k} - e^{-i\phi} \: \BA{j} \BC{k} ) \Big] } $ 
        \\ \hline
    \end{tabular}
    
    \caption{List of some common Gaussian unitaries}
    
    \label{tab: CVgates}
\end{table}


Next, consider the so-called position and momentum 
quadrature operators: 
\begin{equation}
\BQ{} 
= \sqrt{\frac{\hbar}{2}} \: ( \BC{} + \BA{} ), \quad
\BP{} 
= i \: \sqrt{\frac{\hbar}{2}} \: ( \BC{} - \BA{} ).   
\end{equation}
Here, $\BQ{}$ and $\BP{}$ are the position and momentum operators, respectively.
The quadrature operators are observables with continuous spectra
\begin{equation}
\BQ{} \ket{q}
= q \ket{q}, \quad
\BP{} \ket{p}
= p \ket{p},    
\end{equation}
where $ \{ \ket{q} | \: q \in \mathbb{R} \} $ and 
$ \{ \ket{p} | \: p \in \mathbb{R} \} $ are two possible continuous variable (CV) bases for a single qumode Hilbert space. 
The transformation between the position and momentum bases is given by the Fourier transform \cite{Weedbrook2012}
\begin{equation}
\ket{q}
= \frac{1}{\sqrt{2\pi}} \int dp \: e^{- i q p / 2} \ket{p}, \quad
\ket{p}
= \frac{1}{\sqrt{2\pi}} \int dq \: e^{i q p / 2} \ket{q}. 
\end{equation}

The existence of position and momentum CV bases also makes it possible to represent quantum states in terms of wave functions $\langle q | \psi \rangle = \psi (q)$ and $\langle p | \psi \rangle =\tilde{\psi} (p)$, in the position or momentum representations, respectively. 
Of particular importance for the bosonic devices under consideration in this paper are Gaussian states, a class of CV states that can be represented in terms of Gaussian functions. \cite{Weedbrook2012}
One such Gaussian state is the ground state (also known as the vacuum state) of a single qumode, $\langle q | 0 \rangle = \psi_0 (q)$. 
Thus, Gaussian transformations such as phase space displacement and squeezing operators can be defined as operators that map Gaussian states to Gaussian states. \cite{Weedbrook2012}
Some of the Gaussian unitary operators are listed in \Table{\ref{tab: CVgates}}.
However, for universal quantum computations, the Gaussian operations have to be combined with non-Gaussian unitaries. \cite{Braunstein2005quantum,BosonicQiskit2022}
The implementation of non-Gaussian operations can be efficiently done on the bosonic circuit quantum electrodynamics (cQED) devices, as discussed below.

\subsection{Hardware}
\label{Sec:Hardware}

The construction of a quantum device out of a set of qumodes must follow the points below.
First, the qumodes must be able to store information long enough for the computational task to be completed. 
Second, the quantum states of QHOs must be prepared, evolved in time according to a chosen unitary transformation, and measured by controlling with an external system without introducing additional decoherence.
Within the cQED architecture, a set of superconducting resonators act as qumodes, and a non-linear auxiliary circuit is used to perform those three functions (initial preparation, unitary evolution, and measurement).
\cite{Deleglise2008,Hacker2019,Copetudo2024}

Each resonator can be thought of as a quantum LC circuit, which is a quantized version of a classical LC circuit consisting of an inductor (coil) of inductance $L$ and a capacitor of capacitance $C$. 
A classical LC circuit behaves as a mechanical system that undergoes simple harmonic motion, such as a mass on a spring. 
In the circuit, electric charge and current fulfill the role, respectively, of position and velocity in a mechanical spring. The current moves back and forth between the inductor and the capacitor with a resonant frequency given by 
$ \omega = 1 / \sqrt{LC} $. 
This is also the frequency of the quantum LC circuit which realizes a quantum harmonic oscillator.
The resonator geometry design critically affects the performance of the quantum devices. 
Early resonator prototypes focused on the planar design because of the convenience of fabricating superconducting metal circuits on the same chip that contains the auxiliary circuits with standard lithographic tools. \cite{Megrant2012planar}
A range of geometries have been introduced since, including 3D resonators that can produce coherence time up to milliseconds by taking advantage of their hollow structure. \cite{Copetudo2024}
The coupling to the auxiliary circuit can now be achieved, for example, by introducing a center pin in a cylindrical resonator, while placing the auxiliary circuit next to the resonator. \cite{Reagor2016}

The auxiliary circuits that are used to couple the qumodes are made of a Josephson junction, a device of two layers of superconducting material separated by a thin insulator. 
In a superconducting circuit, a Josephson junction substitutes the inductor of the LC circuit and the current is generated by Cooper pairs (pairs of bound electrons) that tunnel through the junction. \cite{devoret2004superconducting,siddiqi2021engineering}
The Josephson junction becomes insensitive to charge noise in the regime of large Josephson energy to the charging energy ratio, \cite{Koch2007}
which is commonly referred to as a \textit{transmon}. \cite{Blais2021,Roth2023}
The transmon superconducting circuit works as an anharmonic oscillator where the levels are no longer equidistant, which allows for selecting the transitions between energy levels. 
The two lowest levels are used in the transmon qubits, which are the predominant element in circuit-based quantum hardware, such as the IBM and Google quantum processors. \cite{steffen2011quantum, google2020hartree}

The transmon is coupled to the qumode resonator by extending its superconducting layers to overlap with the resonator’s electromagnetic field. 
The interaction between the auxiliary circuits and qumodes can be controlled by detuning the qumode and transmon resonance frequencies or changing the overlap between their electromagnetic fields and make it possible to implement the non-Gaussian operations in cQED. \cite{Copetudo2024}
The non-linearities of the cQED devices underlying the non-Gaussian operations are also an exciting avenue to explore. 
Anharmonic resonators, such as superconducting nonlinear asymmetric inductive element (SNAIL) \cite{Sivak2019SNAIL} will be discussed in more detail in later sections of this perspective.


\section{Recent developments} \label{sec: past_work}

Some recent applications of bosonic quantum devices to chemical problems are discussed in this section.

\subsection{Molecular vibronic spectra}


\begin{figure}[t]
\centering
\includegraphics[width=0.9\columnwidth]{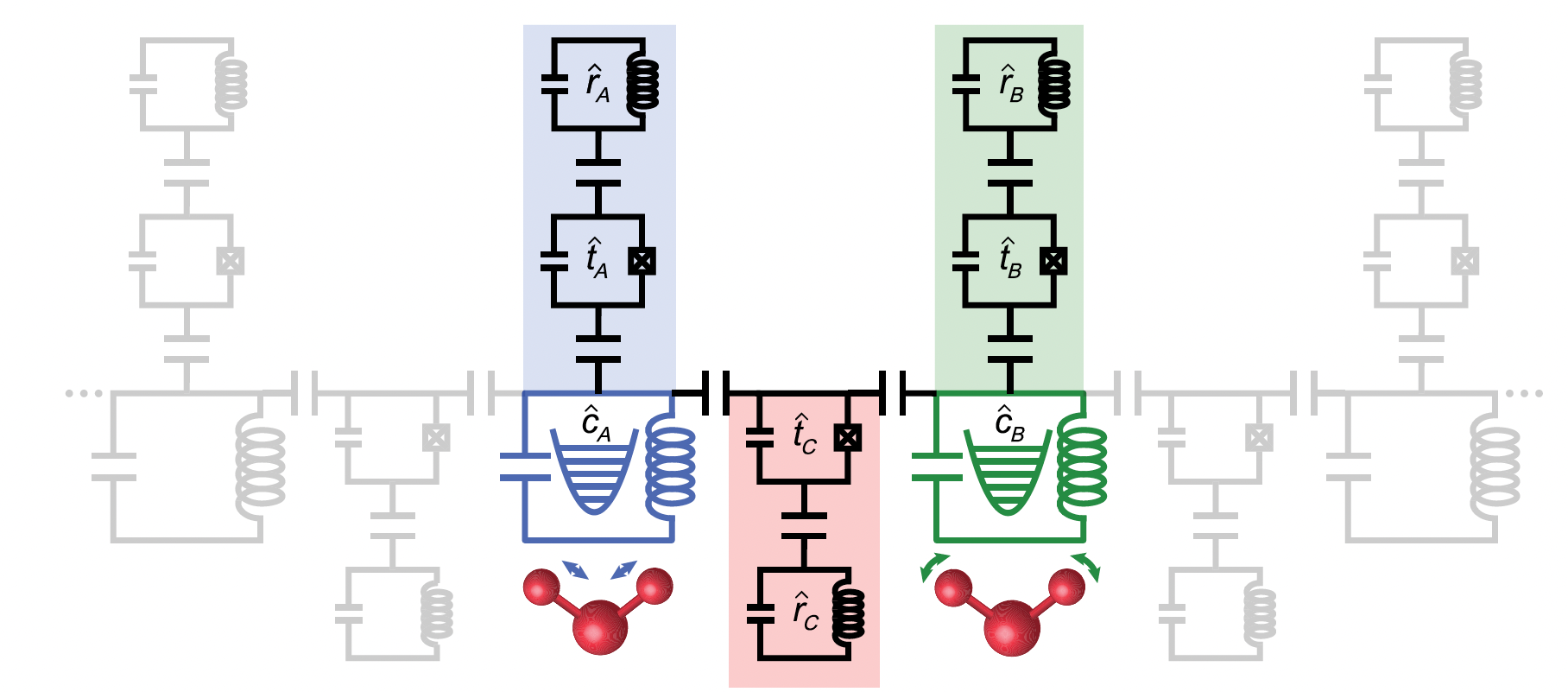}
\caption{
    Schematic of the superconducting bosonic device setup for simulating the vibration of a water molecule. 
    Here, $\hat{c}_A$ and $\hat{c}_B$ are the microwave cavity modes representing the two vibrational modes.
    Transmon qubits $\hat{t}_A, \hat{t}_B$ and readout resonators $\hat{r}_A, \hat{r}_B$ take care of state preparation and measurement. 
    The coupler transmon $\hat{t}_C$ is used for Gaussian operations on the cavity modes and the readout resonator $\hat{r}_c$ is used for characterization and postselection.
    Reprinted from \Reference{\citenum{Wang2020}}.
}
\label{fig: water_circuit}
\end{figure}


Molecular spectroscopy 
is a highly sensitive probe of chemical structure and dynamics. 
\cite{PaviaBook,Barone2021} 
However, the simulation of vibronic spectra, which measures transitions between vibrational states in manifolds that correspond to different electronic states, is a challenging computational task. 
\cite{Dierksen2004}
Specifically, 
predicting molecular vibronic spectra 
under the Born--Oppenheimer approximation
calls for the efficient calculation of Franck--Condon factors (FCFs), which are given in terms of the modulus square of the overlap  between the initial and final vibrational wavefunctions corresponding to their potential energy surfaces.
Recent studies have demonstrated that the calculation of vibronic spectra can be carried out efficiently on bosonic devices. 
This is because the computation of FCFs is related to the boson sampling problem. \cite{Huh2015,Wang2020,quesada2019franck} 

To date, cQED devices have been used to demonstrate the vibronic spectra of H$_2$O, O$_3$, NO$_2$, and SO$_2$ molecules.\cite{Wang2020}
In what follows, we discuss the case of water in some detail, as a specific example. 


\begin{figure}[t]
\centering
\includegraphics[width=0.9\columnwidth]{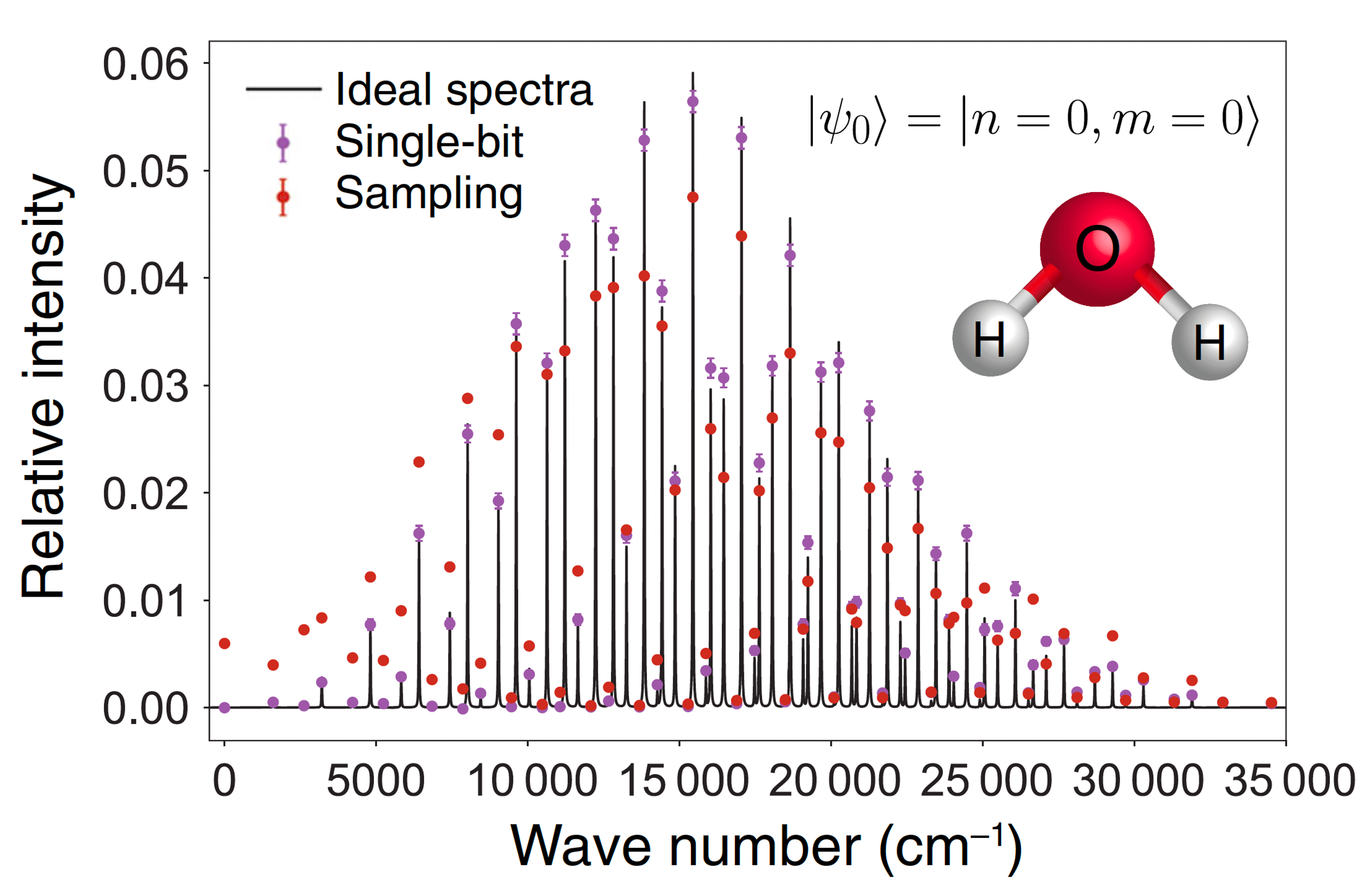}
\caption{
    Vibronic spectra for the photoionization of water to the ($\tilde{B}^2 B_2$) excited state of the H$_2$O$^+$ cation, starting from the vibrationless H$_2$O molecule.
    Reprinted from \Reference{\citenum{Wang2020}}.
}
\label{fig: water_spectra}
\end{figure}


We focus on the symmetric stretching and bending modes of the water molecule here.
Those two vibrational modes are represented by two microwave cavity resonators (i.e., the two qumodes) within the cQED device. 
The local ports in the cavity modes are responsible for displacement operations.
A transmon couples these resonators for beam-splitting and squeezing operations and is connected to a third readout resonator. 
The state preparation and measurement can be done with an additional couple of transmon-readout systems.
We show the schematic of the cQED device for the water molecule in \Fig{\ref{fig: water_circuit}}.

The computation of FCF involves encoding the relation between initial and final normal coordinates corresponding to the potential energy surfaces for the vibronic transitions, which is known as the Duschinsky transformation. \cite{duschinsky1937importance}
In the Hilbert space representation, the Duschinsky transformation can be represented by the so-called Doktorov operator, \cite{doktorov1975dynamical,doktorov1977dynamical} and the FCF for two vibrational modes becomes 
\begin{equation}
FCF_{n, m, n', m'} 
= | \braket{ n, m | \hat{U}_{\text{Dok}} | n', m'} |^2~~.     
\end{equation}
Here, $\ket{n, m}$ is the Fock state corresponding to the initial vibronic state, $\ket{n', m'}$ is the Fock state corresponding to the final vibronic state,
and $\hat{U}_{\text{Dok}}$ is the Doktorov operator representing the Duschinsky transformation in terms of two single-mode displacement, two single-mode squeezing, and one two-mode beam-splitter operations, 
which correspond to the displacement, distortion, and rotation of the potential energy surfaces due to the vibronic transition. \cite{Huh2015,Wang2020,quesada2019franck} 
Thus, after preparing the initial Fock state $\ket{m, n}$ in the cavity resonators, the Doktorov operator is applied followed by a set of measurements. 
In the final step, the photon number measurements are averaged to compute the probability of finding the Fock state $\ket{n', m'}$, which is equivalent to computing the corresponding FCFs. 
The measurements can be done by either single-bit extraction, which maps $\ket{n', m'}$ to a state of two transmons, or by single-shot photon number resolving detection, which measures the number of photons in a given cavity mode. \cite{Wang2020}
The simulated vibronic spectra for the photoionization of the water molecule,
$ \text{H}_2\text{O} \xrightarrow{h \nu} \text{H}_2\text{O}^{+} + e^{-} $ starting from the vibrational state $\ket{0, 0}$ is shown in \Fig{\ref{fig: water_spectra}}.
It has been shown that simulating the same spectra needs 8 qubits and $\mathcal{O} (10^3)$ gates for a qubit-based algorithm, \cite{Wang2020} validating the advantage of bosonic quantum devices for bosonic problems.

\subsection{Vibrational spectra of nonrigid molecules}

The normal mode (harmonic) approximation for molecular vibrations holds when the amplitudes of the oscillations are small. But
there is a broad range of nonrigid molecules for which one or more vibrational modes are anharmonic and the associated nuclear displacements are large. 
For example, such large amplitude modes are described by complicated potentials that have at least two minima. The best known cases are the bending vibration in the water molecule~\cite{Child1999, Zobov2006} and the $\nu_7$ bending mode of cyanogen iso-thiocyanate (NCNCS)~\cite{Winnewisser2005, Tokaryk2023}.

A successful approach to understanding the complex vibrational spectra of nonrigid molecules consists of approximating their energy potential function as discussed in \Reference{\citenum{Child1998}}.
When the excitation energy reaches the local maximum at the origin of the potential, the nature of the vibration switches from bent to linear, and complex features arise in the spectrum. This sudden change in the nature of the excited states of a system as it goes through a critical excitation energy has become known as an excited state quantum phase transition (ESQPT) \cite{Cejnar2021}, in analogy with the concept of a quantum phase transition (QPT), which corresponds to a sudden change in the ground state of the system as a control parameter reaches a critical point. \cite{SachdevBookQPT,Wu2004QPT}
One of the main signatures of an ESQPT is a singularity in the density of states at the critical excitation energy, which moves to higher excitation energies as the control parameter increases.

\begin{figure*}[h!]
	\includegraphics[width=0.8\textwidth]{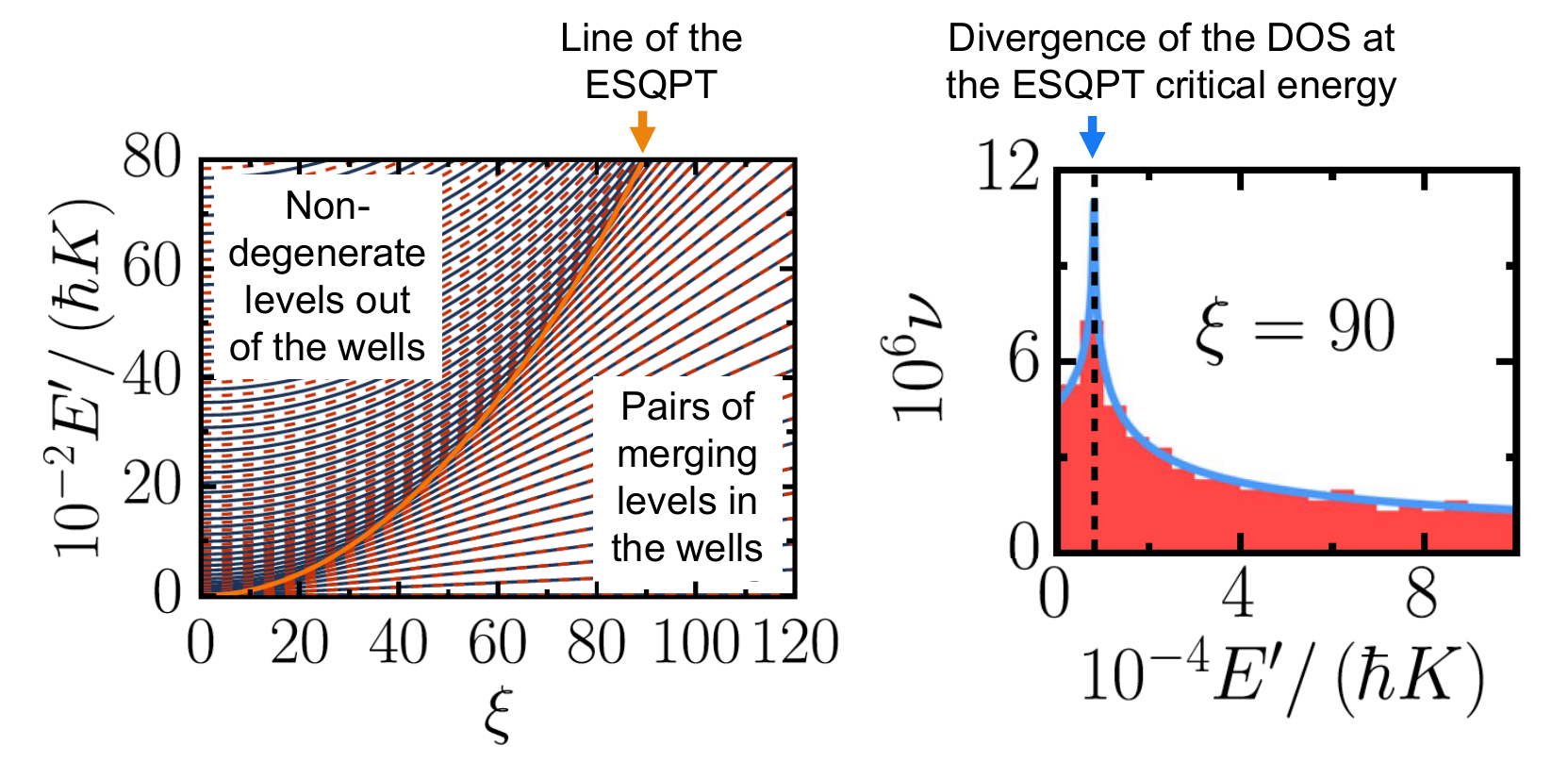}
	\caption{Left: Excitation energies of the Hamiltonian in Eq.~(\ref{Eq:Ham}) as a function of the control parameter $\xi$. 
	Solid black lines are for the levels in the even parity sector, dashed red lines are for the odd parity, and the thick solid orange line marks the energy of the ESQPT. Right:
	Normalized density of states, where the shade indicates numerical results and the solid line is an analytical curve. The vertical dashed line marks the ESQPT energy (more details in \Reference{\citenum{Chavez2023}}).
	}
		\label{figBeSa}
	\end{figure*}

There are numerous systems that exhibit an ESQPT~\cite{Cejnar2021} and the first experimental detection of such transition was done in molecules~\cite{Larese2011,Larese2013,Winnewisser2005,Winnewisser2006,Zobov2006, Khalouf2021,Tokaryk2023}. 
Recently, it has been argued~\cite{Chavez2023} that the superconducting circuit used to generate 
an anharmonic resonator offers an excellent platform to analyze the spectra of nonrigid molecules and ESQPTs in general. This is because the static effective Hamiltonian of the driven anharmonic resonator~\cite{Frattini2022},
\begin{equation}
\frac{\hat{H}_{qu}}{\hbar\, K} =   \hat{n} (\hat{n}-1) -  \xi \left( \hat{a}^{\dagger 2} +  \hat{a}^2  \right),
\label{Eq:Ham}
\end{equation}
 describes a double-well metapotential that captures the ESQPT. In the equation above, $\hat{n}=\hat{a}^{\dagger} \hat{a}$, $K$ is the Kerr nonlinearity and $\xi$ is the control parameter. We denote the eigenvalues of $\hat{H}_{qu}$ by $E$ and the ground state energy by $E_0$. Figure~\ref{figBeSa} shows the main features of the ESQPT associated with $\hat{H}_{qu}$. The left panel displays the excitation energies, $E'=(E-E_0)$ as a function of the control parameter $\xi$. The orange solid line marks the critical energy of the ESQPT for a given value of the control parameter. Below this line, the levels are within the double-well metapotential and pairs of adjacent eigenvalues (black and red lines) merge together (``spectral kissing'' \cite{Frattini2022}), while above the ESQPT line the levels are outside the wells and are non-degenerate. The right panel shows the peak of the density of states at the ESQPT energy. This peak diverges in the mean-field limit.

The analysis of the spectra of nonrigid molecules with superconducting circuit platforms can be extended to more elaborate scenarios by coupling two or more oscillators. The single-driven nonlinear oscillator examined in Eq.~(\ref{Eq:Ham}) has just one degree of freedom, while a full description of the vibrational bending modes in nonrigid molecules requires two degrees of freedom. By coupling two oscillators, an additional degree of freedom can be accessed, with which molecular spectra including the vibrational angular momentum dependence could be reproduced. A second research line that can be investigated is the use of two coupled qumodes, each one describing a local stretching mode, as a simulator of the transition from normal to local stretching vibrations. In addition to spectroscopy, the superconducting circuit platform discussed here could be used as a simulator for other phenomena related to ESQPTs, such as isomerization reactions \cite{Baraban2015,Khalouf2019}. The concentration of energy levels at the ESQPT critical point is analogous to the approach of the neighboring energy levels that happen close to an isomerization barrier height.
These examples provide new research avenues to explore in the near future.

\subsection{Quantum dynamics}


\begin{figure}[t]
\centering
\includegraphics[width=0.7\columnwidth]{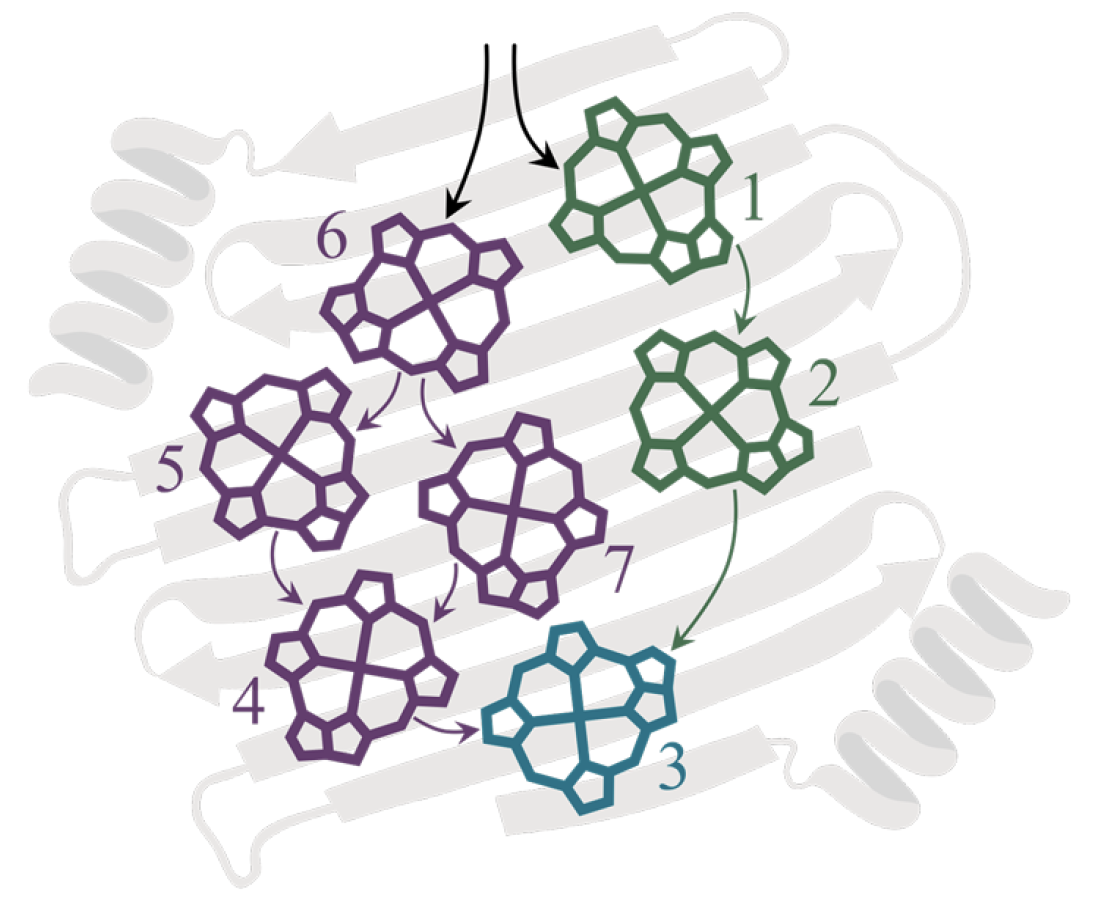}
\caption{
    Schematic for seven chromophores in Fenna--Matthews--Olson (FMO) complex, where the arrows represent energy transfer.
    Reprinted from \Reference{\citenum{Lyu2023}}.
}
\label{fig: fmo}
\end{figure}


Chemical dynamics is inherently quantum, and for a closed system, it is governed by the time-dependent Schr\"odinger equation. Within this equation, the generator of dynamics is the molecular Hamiltonian.   
It has recently been shown that the molecular Hamiltonian can be mapped onto 
a polynomial of bosonic creation and annihilation operators of a {\em single} bosonic mode (SBM). \cite{Lyu2023} 
Thus, using this so-called {\em SBM mapping} allows for the simulation of chemical dynamics on 
a cQED device.

Specifically, the SBM mapping makes it possible to transform an arbitrary $k \times k$ Hamiltonian matrix 
\begin{equation}\label{Hmn}
\hat{H}=\sum_{\alpha=0}^{k-1}\sum_{\alpha'=0}^{k-1}H_{\alpha\alpha'}|{\alpha}\rangle\langle {\alpha'}|,  
\end{equation}
which is represented in terms of the basis set $\{|\alpha\rangle\}$ of choice, into the following 
elementary bosonic operators of a single qumode $(\hat{a}, \hat{a}^\dagger)$, as follows:\cite{Lyu2023}
\begin{equation}
\label{eq: sbm_map}
\begin{split}
\hat{H}_{\text{sbm}}=\sum_{m=0}^{k-1}\sum_{n=0}^{k-1}H_{nm}\hat{P}_{nm}.    
\end{split}
\end{equation}
where
\begin{equation}\label{sbh}
\hat{P}_{nm} \equiv \frac{1}{(k-1)!^{2}}\sqrt{\frac{m!}{n!}}(\hat{a}^\dagger)^n \hat{\Gamma}_k^{k-1} (\hat{a}^\dagger)^{k-1-m}.
\end{equation}
where $\hat{\Gamma}_k=((k-1)-\hat{n})\hat{a}$. 
For example, the Pauli-Z matrix can be represented as 
\begin{equation}
Z 
= \begin{bmatrix}
1 & 0
\\
0 & -1 
\end{bmatrix}  
\mapsto 
1 - 2 \: \BC{} \BA{},
\end{equation}
using the SBM mapping.
Similarly, any $2 \times 2$ Hamiltonian such as the Pauli operators can be mapped onto the Hamiltonian 
of a currently available bosonic quantum device known as driven superconducting nonlinear asymmetric inductive element (SNAIL) \cite{Sivak2019SNAIL,Hillmann2020,Eriksson2024universal}
\begin{equation}
\Ham_{\text{SNAIL}}
= \hbar \omega \: \BC{} \BA{} 
+ g_3 \: ( \BA{} + \BC{} )^3.
\end{equation}
Given that all 1-qubit gates correspond to unitary dynamics generated by a $2 \times 2$ Hamiltonian, they can therefore be generated by such SNAIL devices. 
Furthermore, $4 \times 4$ Hamiltonian matrices, which correspond to 2-qubit gates,  can also be implemented by combining the SNAIL with a combination of Josephson junctions that gives rise to fourth-order terms in the device Hamiltonian. \cite{Lyu2023}

The SBM mapping has also been used to simulate quantum 
dynamics in several chemical systems, including energy transfer dynamics in the Fenna--Matthews--Olson (FMO) light-harvesting complex and charge-transfer (redox) chemical reactions within the spin-boson model.\cite{Lyu2023} 
As an example, we consider the application of SBM mapping to the FMO complex (see \Fig{\ref{fig: fmo}}).
Assuming that the energy transfer in FMO occurs across four chlorophyll molecules, \cite{Abramavicius2011,Hu2018} a model Hamiltonian for FMO can be represented by the following $4 \times 4$ Hermitian matrix (in units of cm$^{-1}$):\cite{Abramavicius2011,Hu2018}
\begin{equation} \label{eq: fmo_ham}
\Ham_{\text{FMO}}
= \begin{bmatrix}
310.0 & -97.9 & 5.5 & -5.8 
\\
-97.9 & 230.0 & 30.1 & 7.3 
\\
5.5 & 30.1 & 0.0 & -58.8 
\\
-5.8 & 7.3 & -58.8 & 180.0 
\end{bmatrix},
\end{equation}
The diagonal elements represent the excitation energies of the chlorophyll, while the off-diagonal elements represent the couplings between them. 


\begin{figure}[t]
\centering
\includegraphics[width=0.8\columnwidth]{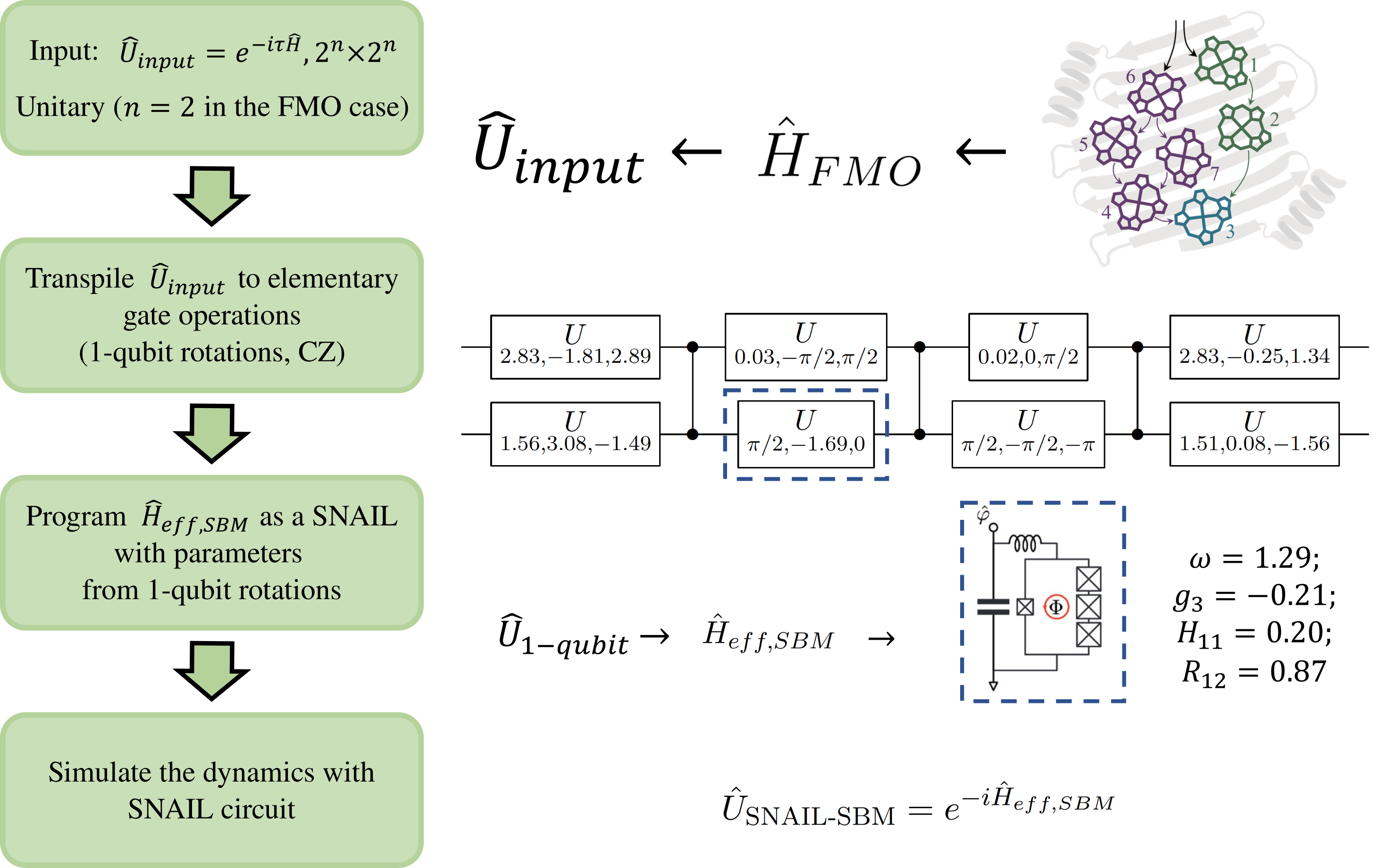}
\caption{
    A schematic protocol for simulating energy transfer in the FMO 4-site model in \Eq{\ref{eq: fmo_ham}} with a SNAIL device. 
    Adapted from \Reference{\citenum{Lyu2023}}.
}
\label{fig: fmo_snail}
\end{figure}


\begin{figure}[t]
\centering
\includegraphics[width=0.8\columnwidth]{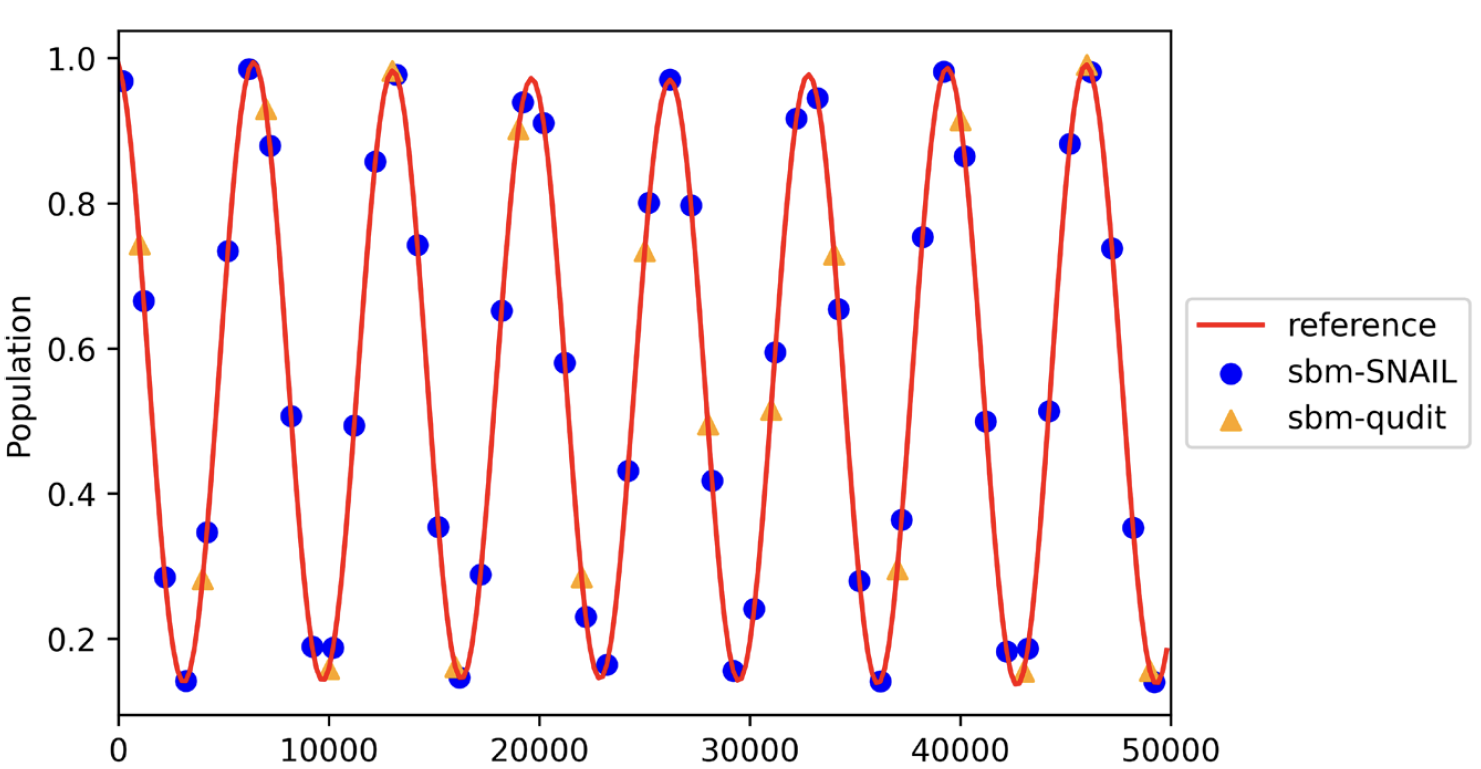}
\caption{
    Comparison between SNAIL simulations with the benchmark data for the population dynamics for the 4-site FMO complex.
    Reprinted from \Reference{\citenum{Lyu2023}}.
}
\label{fig: fmo_plot}
\end{figure}


The dynamics of the 4-site FMO complex, described by the Hamiltonian in \Eq{\ref{eq: fmo_ham}}, can then be simulated with the SBM mapping in conjunction with SNAIL devices. 
This is done by building the unitary propagator corresponding to $\Ham_{\text{FMO}}$ and mapping it onto a SNAIL-based circuit, as shown by the protocol given in \Fig{\ref{fig: fmo_snail}}. 
The SNAIL-based approach to simulating the FMO energy transfer dynamics agrees well with the exact benchmark, as shown in \Fig{\ref{fig: fmo_plot}}.

Since the SBM mapping is a general approach, it can potentially be applied to any quantum dynamics algorithm that has been developed for quantum computers based on qubits. \cite{Ollitrault2021,Wang2023GQME}


Since many systems of chemical interest are coupled to their surrounding environment, the simulation of open quantum systems on quantum computers has recently gained much attention.\cite{hu20, Wang2023GQME, schlimgen21, schlimgen22, schlimgen22b, li24,walters24,seneviratne24} 
Most of these simulations are based on the Lindblad equation which relies on the Born--Markov approximation in the system-environment weak coupling limit. \cite{Wang2023GQME,schlimgen21, schlimgen22} 
Developing quantum algorithms for numerically exact quantum dynamics applied to the regime beyond this weak coupling
may involve either doing part of the computations on a classical computer or resorting to more qubits to represent the environmental degrees of freedom. 
For example, to simulate the dynamics of the spin-boson model, the current qubit-based quantum algorithm developed for the formally exact generalized quantum master equation (GQME) approach 
\cite{Shi2003new,Shi2004semiclassical,Kelly2016generalized,Montoya2016approximate, Montoya2017approximate,Pfalzgraff2019efficient, Mulvihill2019combining,Mulvihill2019modified,Ng2021nonuniqueness, Mulvihill2021roadmap,Mulvihill2021simulating,Mulvihill2022,Lyu2023Tensor, Sayer2024efficient}
relies on a classical computer to solve the underlying exact memory kernel due to the limited computational resources of current NISQ devices.
\cite{Wang2023GQME}
The hierarchical equations of motion (HEOM) approach is a numerically exact method for open quantum systems,\cite{tanimura89,tanimura06,Tanimura2020,shi09c,shi18,dan23,dan23b} which decomposes the environment into a few dissipative bosonic modes, \cite{ke22,li24} and as a result needs extra qubits for encoding. 
Bosonic quantum devices provide a natural way to encode the dissipative modes with qumodes to simulate the dynamics of quantum systems coupled to a bosonic environment.

\section{Ongoing directions} \label{sec: current_work}

In this section, several new frontiers for simulating chemistry on bosonic quantum devices are outlined.


\subsection{Elementary chemical reaction dynamics}


\begin{figure}[ht]
\centering
\includegraphics[width=0.8\columnwidth]{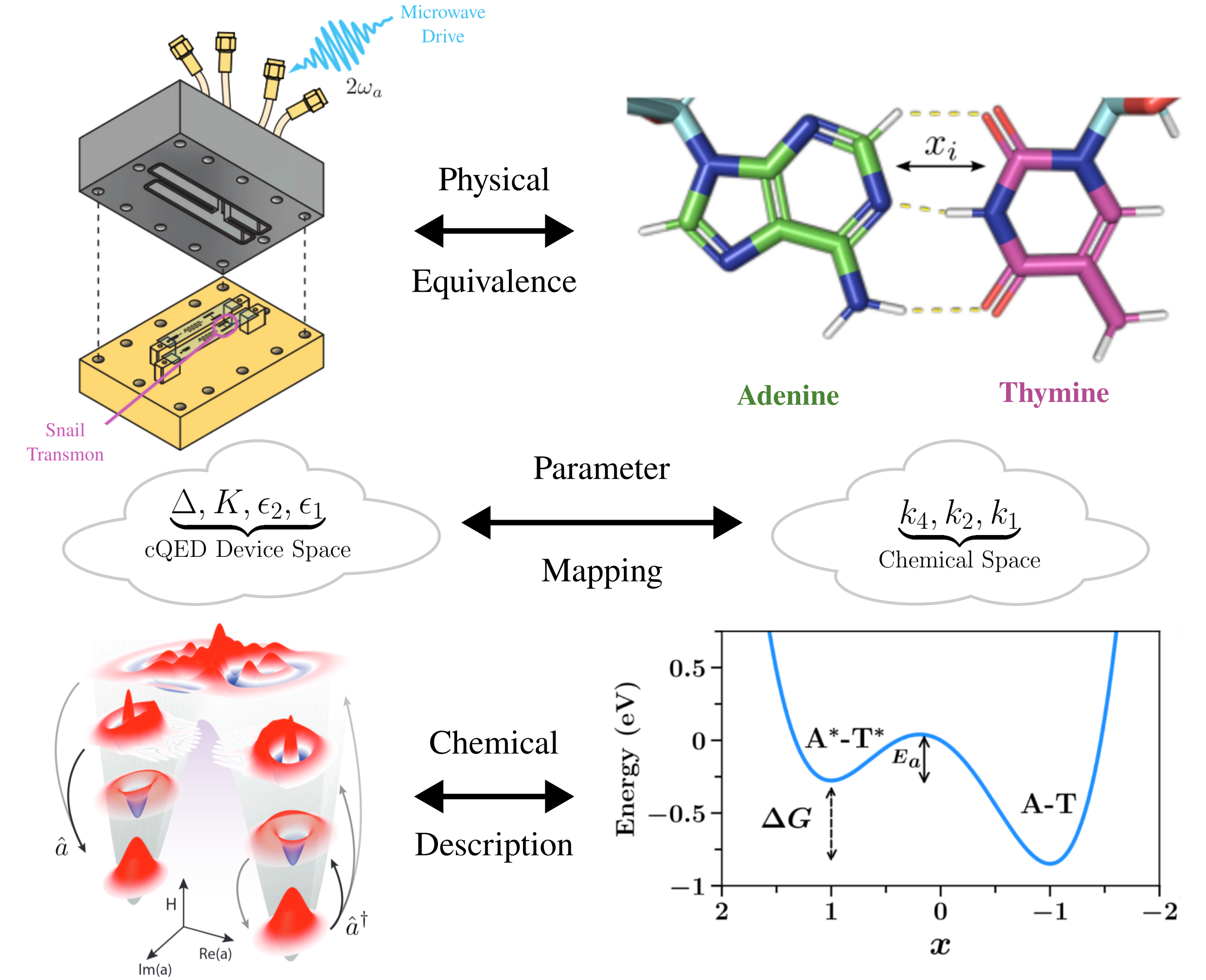}
\caption{
    The correspondence between the bosonic cQED platform device and the chemical systems via a double-well isomorphism. 
    The bosonic cQED device (top left corner, adapted from \Reference{\citenum{Frattini2022}}) can be used to perform analog simulation of chemistry (top right), according to the mapping between the Hamiltonian parameter space of the Kerr-Cat Hamiltonian (middle left) and a generic chemical double-well (middle right).
    The bottom right figure showcases state functions of the device Hamiltonian plotted in a phase space representation (adapted from \Reference{\citenum{Frattini2022}}), which can be thought of as analogous to the states associated with a chemical double-well potential.
}
\label{fig: kerr_cat_to_chem}
\end{figure}


In this section, we give an outlook regarding the asymmetric Kerr-cat as it relates to chemistry, originally from \Reference{\citenum{Albornoz2024}} and \Reference{\citenum{Allen2024}}. 
Bosonic quantum platforms such as the cQED-based Kerr-cat devices offer a general strategy for exploring the chemical dynamics of reactions that can be described by a double-well potential energy surface. \cite{Frattini2022}
The cQED platforms allow tuning dissipation for accurate modeling of chemical reactivity in dissipative environments.
Given that chemical dynamics is inherently quantum-mechanical in nature, 
simulating chemical dynamics on bosonic devices provides 
a novel avenue for exploring chemical reactivity, complementary to traditional approaches based on spectroscopic methods and numerical simulations.

The adenine-thymine proton transfer reaction \cite{Florian1994proton} is an example 
of a biologically significant chemical reaction that can be modeled in terms of a double-well free energy profile, \cite{Godbeer2015modelling}  
as shown in \Fig{\ref{fig: kerr_cat_to_chem}}. 
Simulating the chemical dynamics of the adenine-thymine proton transfer covers a wide range of double-well free energy profiles since it changes significantly during the cell life cycle. 
Indeed, the system can be modeled as a chemical double-well potential ($\hat{U} = k4 \hat{x}^4 - k_2 \hat{x}^2 + k_1 \hat{x}$), where each of the parameters $k_4, k_2, k_1$ are related to the energy landscape from reactants to products, as shown in \Fig{\ref{fig: kerr_cat_to_chem}}.

The quantum dynamics of the bosonic Kerr-cat cQED device described here have been extensively studied. \cite{Reynoso2023, Chavez2023, Iachello2023, Gonzalez2024} 
This experimental platform can be thought of as a double-well analog, as shown in \Fig{\ref{fig: kerr_cat_to_chem}}, with kinetic behavior that depends on the drives of the system as well as the inherent dissipation of the physical experiment. \cite{Frattini2022,Venkatraman2023}.
The parameters of the device can be continuously tuned, something that is difficult to achieve in an experimental chemical setting due to the correlated chemical properties of different functional groups and rich chemical microenvironment.
Furthermore, given the mapping between the device and 
chemical double-well Hamiltonians, this cQED platform offers the
possibility to study a wide range of chemical systems,
and perform high-throughput chemical kinetic studies without the limitations of synthetic chemistry or the numerical limitations of exact open quantum dynamics methods.\cite{Albornoz2024,Allen2024} This is of greater relevance
for chemical problems for which multiple molecular degrees of freedom
are relevant for reactivity for which coupled cQED devices could
address the numerical limitations of high-dimensionality
open quantum dynamics methods.
Thus, the flexibility and the presence of inherent quantum effects in cQED devices can extend and complement the toolset available for studying chemical reaction dynamics, while alleviating the difficulty and labor-intensive efforts of physical chemistry experiments.


\subsection{Photochemical reaction dynamics and conical intersections}


\begin{figure}[t]
\centering
\includegraphics[width=0.7\columnwidth]{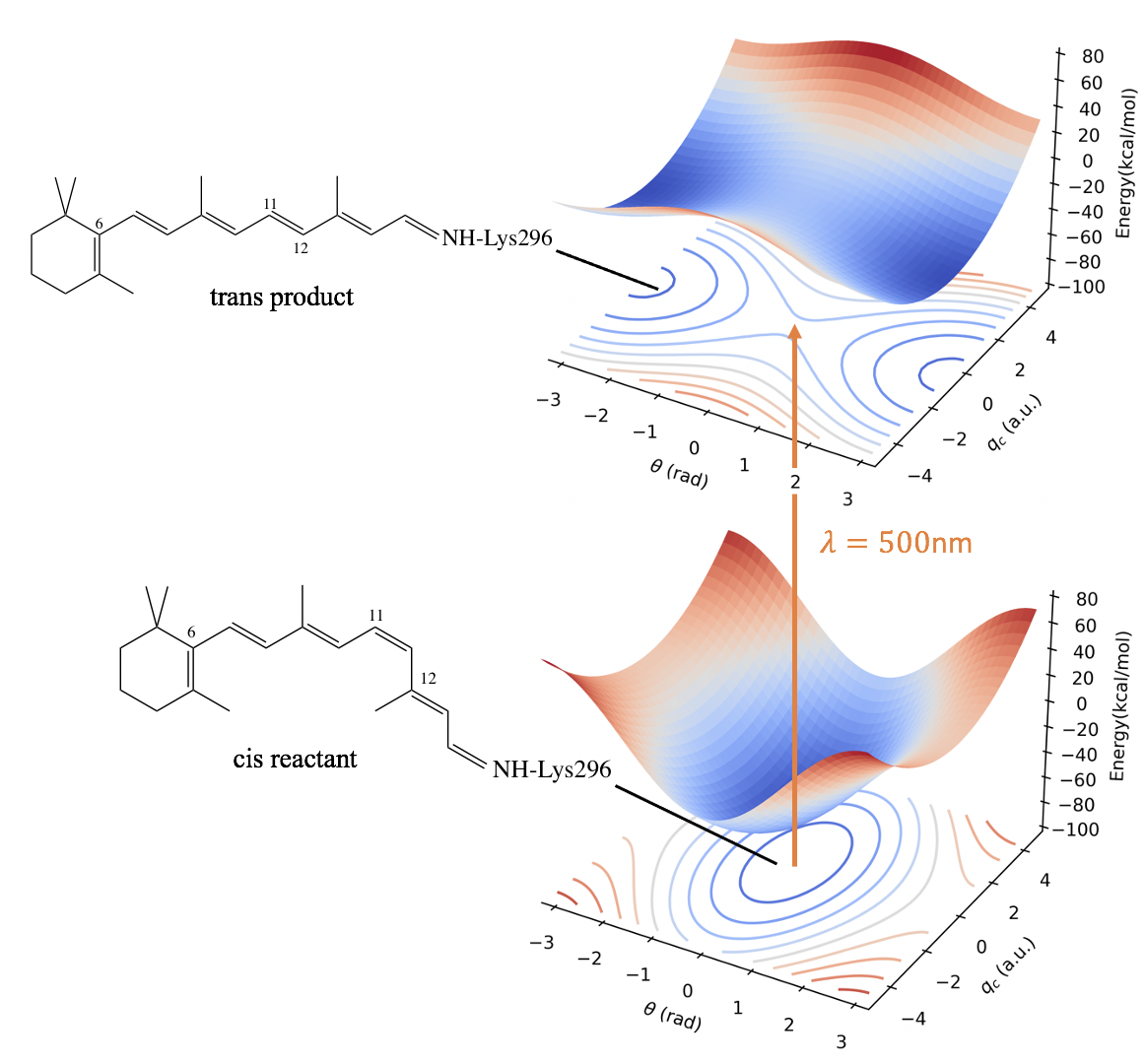}
\caption{
    Two-state, two-dimensional potential energy surface that describes photoinduced isomerization of retinal. 
    Reprinted from \Reference{\citenum{Lyu2022}}. 
}
\label{fig: retinal}
\end{figure}


\begin{figure}[ht]
    \centering
    \includegraphics[width=0.7\columnwidth]{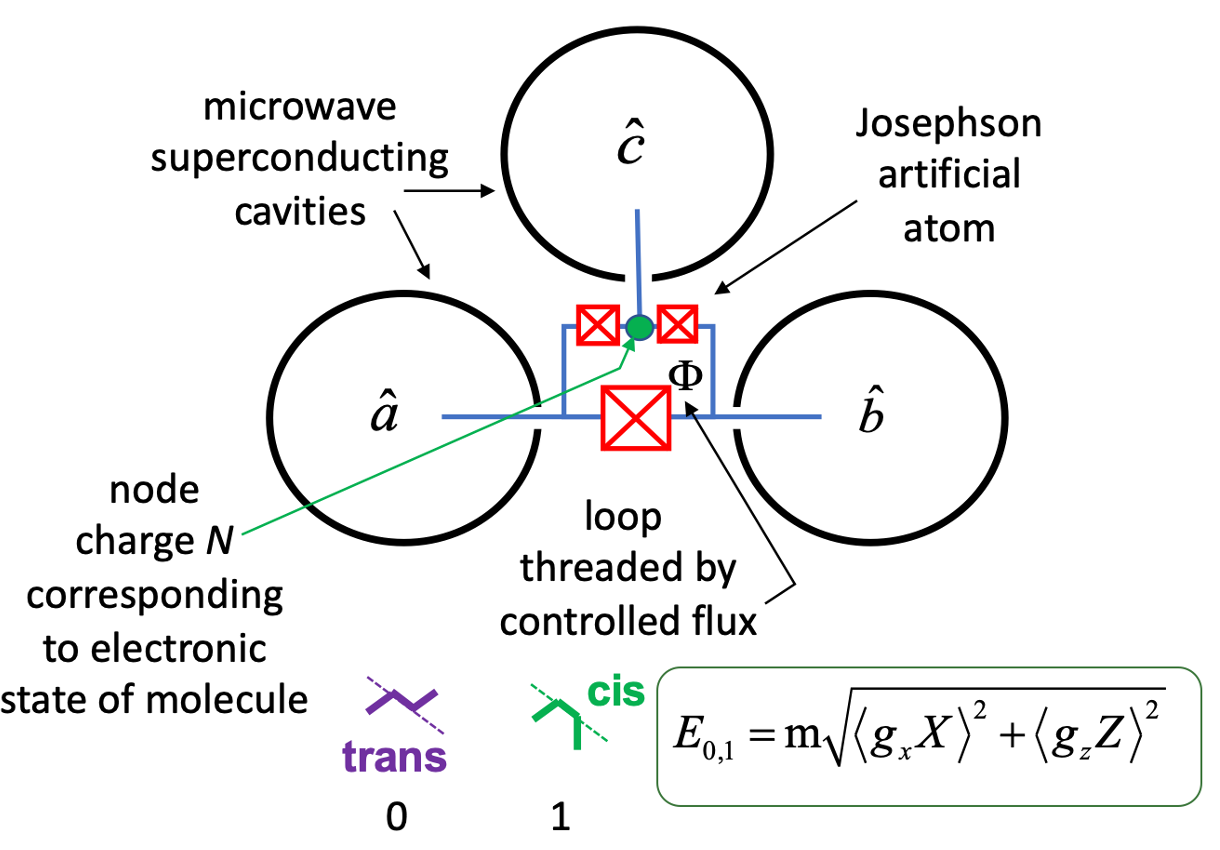}
    \caption{ 
        A proposed cQED circuit for the simulation of a conical intersection between two molecular energy surfaces such as in \Fig{\ref{fig: retinal}}. 
        The artificial chemical system is realizable by a superconducting circuit,
        comprised of three Josephson junctions (indicated by red boxes) \cite{Vion2002}
        describing the electronic properties of the molecules, whereas the green charged node 
        describes the electronic state of the system. 
        The quantronium circuit is coupled to a set of three microwave superconducting cavities,
        encoding the vibrational degrees of freedom of the molecules. 
        The system dynamics are driven by pulses that modulate the excitation of the electronic and vibrational modes.
        The coupling between the quantronium and cavities can act as a surrogate for the vibronic
        coupling of the chemical system around the conical intersection.
    }
    \label{fig: ci_device}
\end{figure}


Conical intersections are formed when electronic states corresponding to different molecular geometries become energetically degenerate and are critical for simulating photochemical reactions. 
 \cite{Matsika2007,Domcke2012}
Recently, an experimental demonstration of an engineered conical intersection in a hybrid cQED processor for a minimal model of a two-dimensional linear vibronic coupling has shown promise in the use of bosonic quantum devices as tunable simulators of chemical reactivity. \cite{Wang2023}
For the model considered, experimental measurements captured the delicate interplay between coherent evolution and dissipation that is ubiquitous in chemical reactivity. 
By engineering a cQED device with five simultaneous microwave drives and tunable dissipation and performing Wigner state tomography, \textit{branching} of the wavepacket was observed when dissipative coupling induced electronic excitation corresponding to \textit{jumps} between potential energy surfaces. \cite{Wang2023}

The success in engineering the conical intersection with a cavity-transmon device motivates further studies on exploring nonadiabatic chemical dynamics through conical intersections with a bosonic cQED machine. 
For an initial exploration, a model system that describes the ultrafast photoisomerization of retinal molecules might be most suitable. \cite{Lyu2022}
This two-surface, two-dimensional model system describes the $S_0$ and $S_1$ electronic surfaces, each containing two coordinates, namely, the isomerization dihedral angle reaction coordinate and a polyene stretching coordinate, as shown in \Fig{\ref{fig: retinal}}. 
The coupling between the two surfaces is linear in the stretching coordinate, drawing proximity with the model system in \Reference{\citenum{Wang2023}}. 
However, different from that system, the retinal reactive coordinate is modeled with a cosine function for describing the cis-trans isomerization and is therefore highly anharmonic. \cite{Lyu2022}
It is possible to represent the retinal model system with a Hamiltonian of individual transmons, that encode the electronic states, and a set of microwave cavities that encode vibrational degrees of freedom. 
The anharmonicity makes this an example of a numerically challenging problem, and thus obtaining an exact open quantum dynamics description both at short and long times could give insight into the photophysics of the problem. Thus, a transmon-based device could provide a bridge between state-of-the-art numerical methods and existing photochemical experiments, by explicit inclusion of term-by-term quantum effects that 
inherently incorporate the entanglement of multiple potential energy surfaces and other contributing environmental degrees of freedom.
Thus, we envision the retinal potential energy surface to be effectively represented with a transmon-based device such as the one shown in \Fig{\ref{fig: ci_device}}, \cite{Vion2002} 
and serve
as a proof of concept for analogous problems that may be challenging
to exactly simulate on a classical device due to the dimensionality of the problem.

Hence, we demonstrate how this platform can facilitate the investigation of chemical dynamics around conical intersections that harness the capabilities of the bosonic cQED machines.

\subsection{Molecular graph problems}


\begin{figure}[ht]
    \centering
    \includegraphics[scale=0.9]{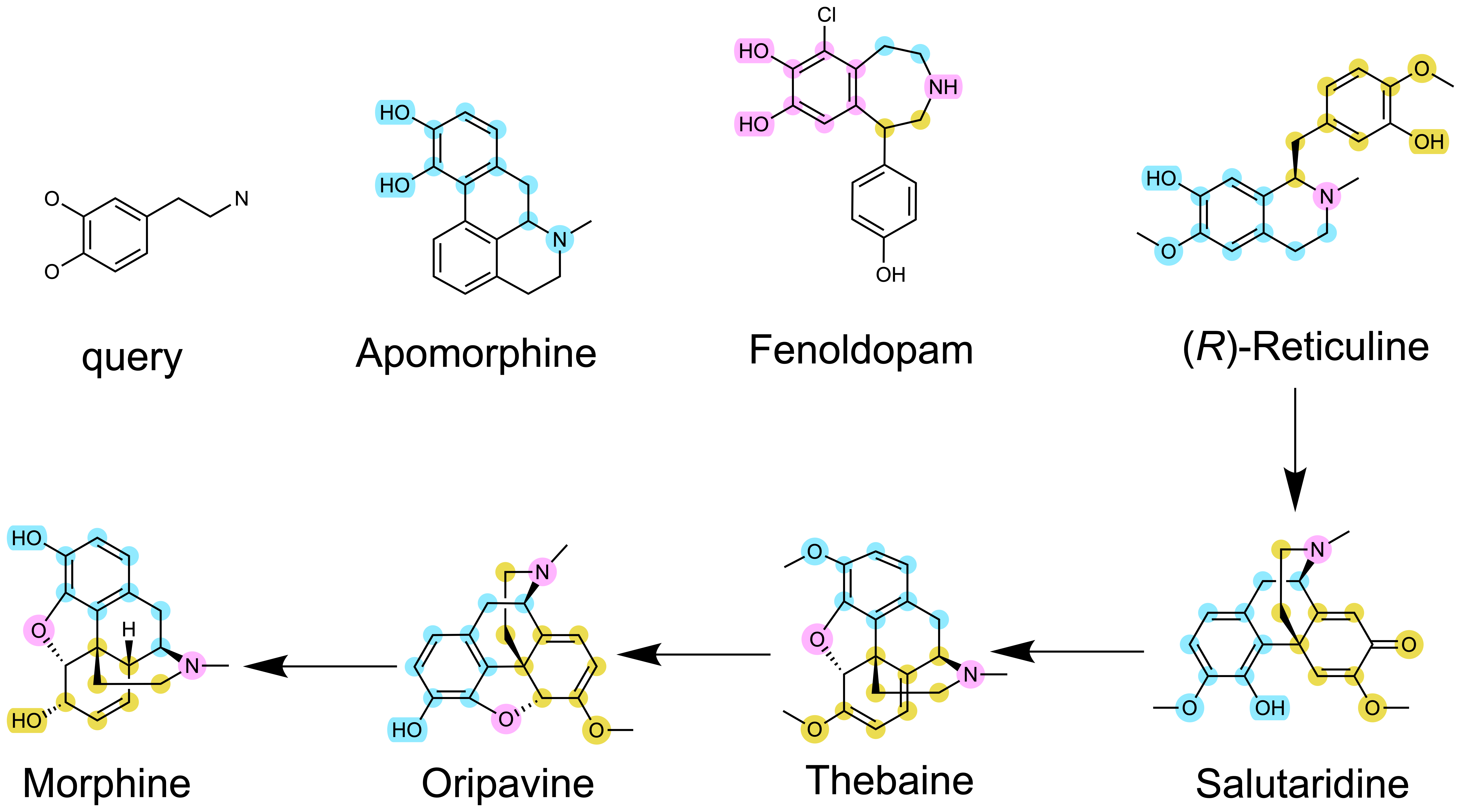}
    \caption{
        Molecular substructure search of compounds that contain the query structure. 
        Apomorphine only admits one such substructure, whereas others contain at least two. 
        The first and second substructure's atoms are in blue and yellow, respectively, where the overlapping atoms are in pink. 
        It 
        has been observed that the biosynthesis of morphine in opium poppy, \cite{Novak2000} starting from \textit{(R)}-reticuline through three other intermediates, preserves the two substructures (not necessarily their conformations) that are similar to the query.
       }
    \label{fig: molecular_graphs}
\end{figure}


The advent of 
big data has led to the mass digitizing of information, including molecular data relevant to chemistry and biosciences. 
Naturally represented by graphs, molecules (and thus chemical reactions) can be stored electronically as massive databases, such as PubChem \cite{Kim2023-fc}, SciFinder, RCSB Protein Data Bank \cite{RCSB}, just to name a few. 
It should be of no surprise that solving graph problems provides quantitative insights into our knowledge of the chemical properties of molecules, proteins, and macromolecules. 
Many graph problems, however, are known to be computationally challenging on larger scales. 
For instance, molecular substructure search and retrosynthetic analysis rely on efficiently finding common subgraphs, which is closely related to the subgraph isomorphism problem, as shown in \Fig{\ref{fig: molecular_graphs}}. 
However, solving the subgraph isomorphism problem, 
a nondeterministic polynomial-time (NP) problem~\cite{Cook2023}, 
generally relies on efficient heuristics to speed up existing algorithms, reduction of subgraph isomorphism tests, or relaxing isomorphism conditions~\cite{Bonnici2013SubgraphBiochemData} that is inherently hard on classical computers~\cite{Garey1979computers}, especially for large graphs.

One promising direction to tackle this challenge is to harness the power of photonic quantum computing, which has been demonstrated to gain a quantum advantage in the Gaussian boson sampling (GBS) problem. \cite{GBSQAdvantage}

Pioneered over a decade ago, \cite{Aaronson2013BS} boson sampling (BS) was proposed as a
non-universal quantum computation model
to compute the permanent of matrices which is a \#
P-hard problem.
One photonic implementation
of BS is GBS, which
samples from squeezed input Gaussian states, a non-classical resource, after being passed 
through linear interferometers. 
In doing so, the GBS protocol approximates the hafnians of matrices, related to the number of perfect matchings of graphs,\cite{Hamilton2017GBS, quesada2019franck}
rather than permanents obtained from BS.
Hence, GBS has the potential for solving important molecular graph problems and molecular docking, efficiently simulating phenomena intractable for classical computers.\cite{banchiMolecularDockingGaussian2020}
Unlike other types of quantum computers which struggle with high degrees of entanglement, photonic circuits are capable of creating large quantum systems with millions of entangled modes via time-domain multiplexing techniques where the number of optical devices used is independent of the number of qumodes. \cite{GBSQAdvantage} 
Whereas graph-like structures are ubiquitous in quantum information and for generating or manipulating entanglement, \cite{krastanovHeterogeneousMultipartiteEntanglement2021,heinEntanglementGraphStates2006} encoding graph characteristics into physical modes allows for efficient analog simulation of complex high-dimensional properties in these bosonic quantum systems. 
Exploring the power of GBS beyond known graph problems, i.e., towards optimization \cite{arrazolaQuantumApproximateOptimization2018} also represents fundamental directions, as well as exploring well-known heuristics for graph problems in the quantum setting. Understanding or mitigating noise \cite{head-marsdenQuantumInformationAlgorithms2021} due to photon loss or other factors in these sampling procedures is also critical for realizing advantages.

To apply the state-of-the-art bosonic cQED quantum platform to solve important molecular graph problems with GBS, a \textit{reuse-and-repurpose} approach can be taken for efficiently implementing large (a few tens of optical modes) GBS problems with small, experimentally-accessible cQED devices with only a handful of cavity modes. \cite{Lyu2024} 
This scheme was tested with the problem of binding a thiol-containing aryl sulfonamide ligand to the tumor necrosis factor-$\alpha$ converting enzyme receptor, showing that the approach facilitates efficient simulation of large (50-100 modes) GBS problems with cQED devices that contain only a handful of modes. \cite{Lyu2024} 

The realistic implementation of boson sampling (and thus GBS) only provides approximate sampling, due to imperfections such as decoherence, noise, and photon losses. Thus, the complexity of GBS algorithms should also be compared to their approximate classical counterparts. In light of recently proposed classical sampling algorithms~\cite{Oh2024Algorithms} for graph problems, the development of GBS algorithms is even more relevant for the exploration of the power of GBS in general, as well as to encourage a further expansion of the scope of quantum-inspired algorithms.

\subsection{Electronic structure calculations}


\begin{figure}[ht]
\centering
\includegraphics[width=0.9\columnwidth]{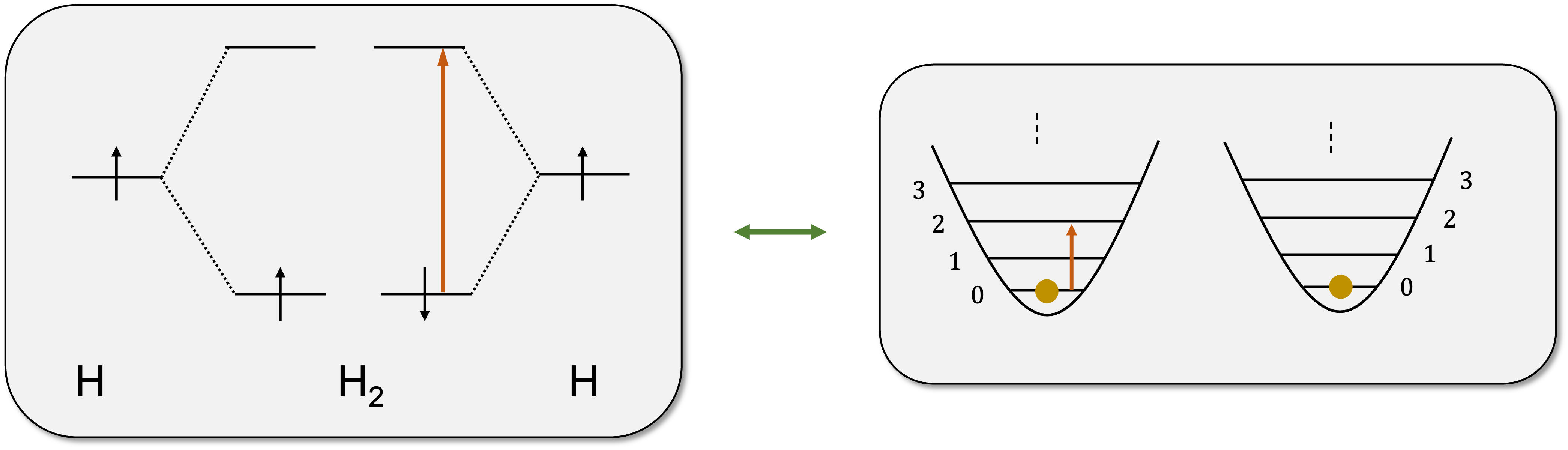}
\caption{
    A schematic of how the electronic states of H$_2$ molecule map to the Fock states of two qumodes. 
    When the two electrons occupy the bonding orbitals of the H$_2$ molecule, that maps to the vacuum state of the two qumodes. 
    The electronic excitations from the occupied orbitals is then mapped to photonic excitations of the qumodes.
}
\label{fig: fbm}
\end{figure}


Bosonic quantum devices are the natural choice for bosonic problems, following the spirit of analog computing. 
However, bosonic quantum devices can achieve universal quantum computation and thus can be applied to any quantum simulation problems. 
One of the most promising applications of quantum computers is the efficient simulation of molecular electronic structure Hamiltonians, which has a far-reaching impact on chemistry and biosciences. \cite{Cao2019,McArdle2020}
Electronic structure simulations provide ground and excited state energies of molecular systems, which are needed to compute reaction rates, understand competing reaction mechanisms, and design new catalysts or life-saving drugs. 

Simulation of a molecular electronic structure on bosonic quantum devices must include a key step of mapping this fermionic problem to a bosonic one.
Traditional quantum computers also tackle a similar problem, where the electronic structure must be mapped to a system of qubits, 
which can be understood as spin-$1/2$ systems or hardcore bosons.
In fact, a Fock space description of qubits reveals that qubits are parafermions: particles with hybrid boson-fermion quantum statistics. \cite{Wu2002qubits}
In the case of qumodes, electronic structure mapping can be achieved by mapping the excitations between molecular orbitals to the excitations between the levels in qumodes. \cite{Ohta1998}
Specifically, a molecular system of $N$ electrons and $M$ spin-orbitals can be mapped to a system of $N$ qumodes with $M - N + 1$ levels. \cite{Dutta2024boson}
As a specific example, we show how electronic states of molecular orbitals, also known as Slater determinants, can be mapped to Fock states of two qumodes for the dihydrogen molecule in \Fig{\ref{fig: fbm}}.
The state mapping can then be used to map the bilinear fermionic operator $E_q^p = f_p^\dagger f_q$ in terms of qumode Fock space projection operators of the form 
$\ket{n_1, \cdots, n_N} \bra{m_1, \cdots, m_N}$, where $f_p^\dagger$ and $f_q$ are fermionic creation and annihilation operators, respectively. \cite{Dhar2006,Dutta2024boson}
The operator mapping is exact and thus preserves the antisymmetry of fermionic algebra. 
The operator mapping only includes a subset of all projection operators possible in the corresponding qumode Fock space, thus the number of terms scales polynomially with system size. 
\cite{Dutta2024boson}
This allows for writing the electronic structure Hamiltonian in the Fock basis of the qumodes, which in turn allows the computation of the molecular electronic energy by a set of Fock number measurements on the cQED device. \cite{Wang2020} 
A bosonic ansatz has also recently been explored for molecular electronic structure combined with fermion to qubit mapping.\cite{shang2024boson}


\subsection{Qudit quantum phase estimation using qumodes}


\begin{figure}[t!]
\centering
\includegraphics[width=0.9\columnwidth]{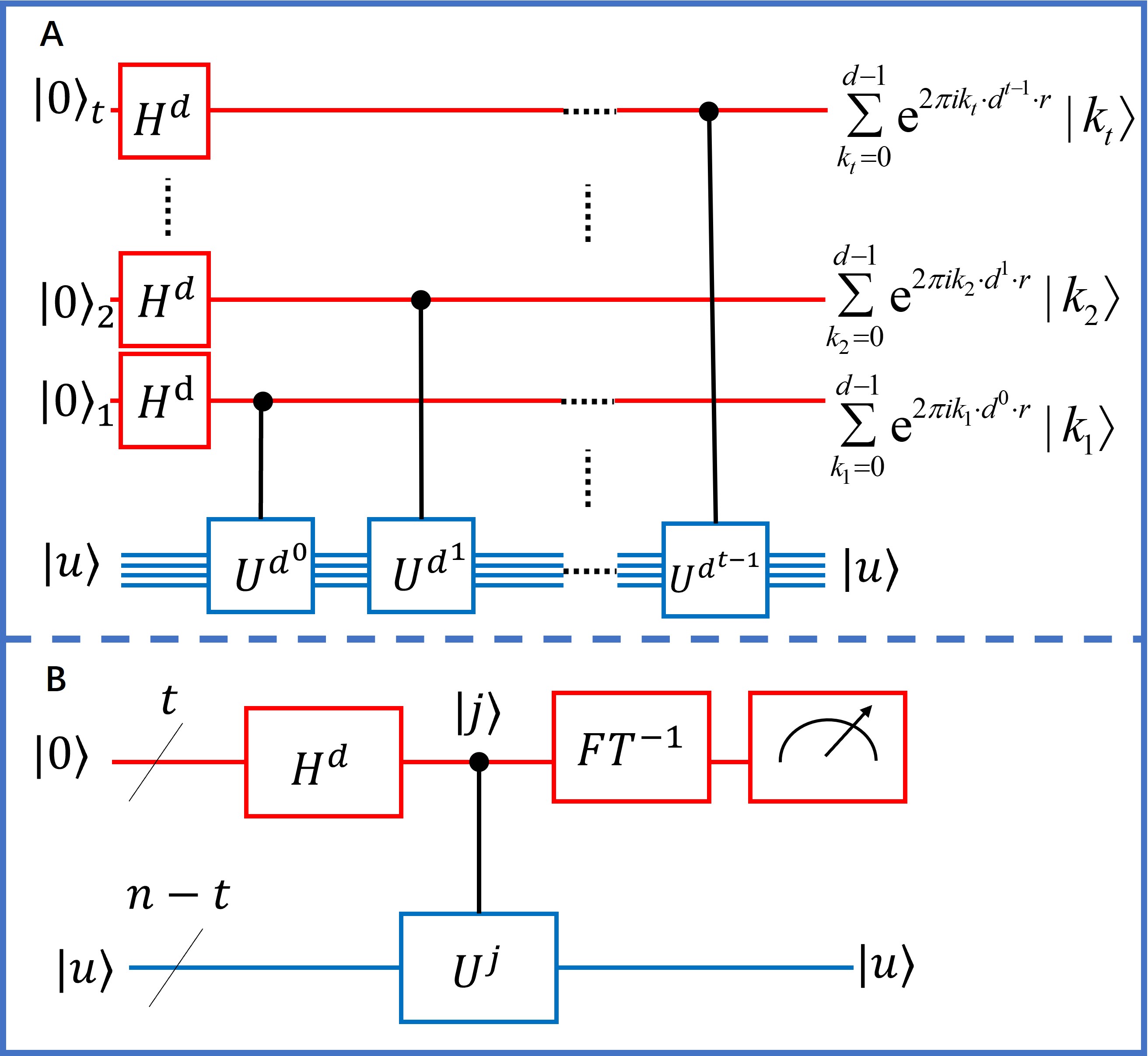}
\caption{
    A) First half of the qudit QPE and the explicit states of the first register qudits after the $U^j$ operations are given on the right. All the qudits used are of dimension $d$. 
    B) The simplified circuit of the qudit QPE. After the Hadamard gate and controlled-$U^j$ operations, the qudit inverse Fourier Transform ($FT^{-1}$) is performed and the phase factor is obtained by measuring the first register qudits states. 
    In both subfigures, the red lines refer to qudits in the first register and the blue lines refer to the qudits in the second register that store the quantum eigenstates of $U$.
}
\label{fig: qudit_PEA}
\end{figure}


Qumodes, with their expanded Hilbert space dimensionality, are natural candidates for representing qudits.
A qudit is a generalization of the qubit in quantum computing. 
While a qubit can exist in states $|0\rangle$ and $|1\rangle$, a qudit can exist in states 
$\{ |0\rangle,|1\rangle,\cdots,|d-1\rangle \} $, where $d$ is the dimension of the qudit. 
So, a qubit is essentially a special case of a qudit where $ d = 2 $. 
Many of the qubit quantum algorithms such as the Quantum Fourier Transform (QFT) and Quantum Phase Estimation (QPE) (as shown in \Fig{\ref{fig: qudit_PEA}}) can be generalized to their qudit versions. \cite{wang2020qudits}
Extending these qubit algorithms to qudits 
reduces the required number of qudits and qudit gates.\cite{Parasa2011}
As the QPE has been applied extensively in quantum chemistry simulations such as electronic structure calculations, the generalization to qudits can benefit these simulations and extend them to more complex chemical systems. 
Generalizing the QPE to the qudit representation provides a notable improvement in terms of reducing the number of required qudits and decreasing the error rate exponentially as the qudit dimension increases~\cite{Parasa2011}. Applications of qudit-based QPE beyond simulating complex chemical systems is a growing area of research. For example, a method to solve the linear system using a qutrit-based QPE algorithm has been proposed~\cite{sawerwain2013quantum}.

While qudits and qumodes represent different aspects of quantum systems—qudits deal with discrete quantum states while qumodes deal with CV quantum states—one could encode qudit states into qumodes.  
This is done by encoding these states into different coherent states of the electromagnetic field. \cite{Optic-Book}
Specifically, we could use $d$ different coherent states, each corresponding to one of the qudit states. 
A 2-qutrit QPE has been demonstrated with a procedure developed for designing the circuit for qudit QPE. \cite{Hsuan-Hao2019,wang2020qudits}
Encoding qudit states into qumodes allows us to implement more qudits with higher dimensions and thus expand the overall computational space.
Implementing the qudit quantum phase estimation algorithm using qumodes can be used for obtaining the eigenvalues of a given Hamiltonian 
such as the potential energy surfaces of molecules.


\section{Concluding remarks} \label{sec: conclusions}

Recent advances in quantum engineering have led to the emergence of modular, highly controllable, and tunable quantum devices which offer a new way to explore chemistry via a \textit{simulate quantum with quantum} approach. \cite{Feynman2018simulating}
While much of the work on the simulation of chemistry on quantum devices has been focused on qubit-based devices, the use of qumode-based devices offers a highly promising alternative approach that can overcome many of the limitations facing currently available qubit-based devices. 
Indeed, the development of qumode-based hardware has seen rapid growth in recent years, giving rise to highly controllable and tunable devices based on combining microwave cavity and superconducting circuit technology.
Matching this progress with similar progress on the application of those devices to simulate chemistry is therefore highly desirable and timely. 

Qumode-based platforms are natural choices for chemical problems that are bosonic, in the spirit of analog computation. 
This is particularly true for problems such as the calculation of molecular vibronic spectra or chemical dynamics on harmonic potential energy surfaces. 
However, efficient mapping schemes can expand the capabilities of bosonic quantum devices to a broader range of chemical problems beyond inherently bosonic problems, as shown by the SBM mapping scheme for quantum dynamics.
Furthermore, despite fermions and bosons representing two different building blocks of matter, fermionic states can be exactly mapped onto bosonic states, which opens the door for performing electronic structure calculations on bosonic quantum devices.

Despite the advantages of the described bosonic cQED platforms, some of the outstanding challenges include increasing the number of coupled qumodes and the number of states that can be resolved within each qumode.\cite{Wang2020} 
The latter point is important for efficiently and accurately encoding chemical problems, such as those involved in dynamics and electronic structure problems, while the former is necessary for answering higher-dimensional chemical questions.
One approach has been to utilize universal bosonic circuits to encode algorithmic operations.
For example, there has been recent progress in developing universal bosonic circuits based on the cQED platform, that can in principle represent any unitary of arbitrary dimension, including gates based on multiple qubits. This can open the door for quantum algorithms that can use the multiple levels of one oscillator rather than using multiple qubits. \cite{eickbusch2022fast, you2024crosstalk, Eriksson2024universal, Hillmann2020}
Thus, we can encode and study chemical problems within a compact bosonic framework, aiding to the development of enhanced, high-quantum-volume bosonic devices with scalable error-tolerance.

Fault-tolerant quantum computing (FTQC), based on quantum error correction (QEC), is the only reliable and credible path toward scalable large-scale general-purpose applications. \cite{Campbell2017roads,Babbush2021focus} 
However, FTQC requires substantial qubit and gate resources. \cite{Babbush2021focus}
Open-loop quantum control methods such as dynamical decoupling (DD) \cite{Viola1999} and its many variants are thus attractive in this regard, as has been demonstrated in recent experiments using transmons. \cite{Pokharel2018demonstration,Tripathi2022suppression,Ezzell2023dynamical,Pokharel2023demonstration, Pokharel2024better,Singkanipa2024demonstration}
Under appropriate conditions, integrating DD with a QEC scheme can lead to a reduced accuracy threshold for FTQC. \cite{Ng2011DD}
In practice, a DD-protected gate (that is, a gate executed following a DD pulse) can achieve greater accuracy than an unprotected gate. A critical challenge remains: to validate DD methods on cQED platforms by demonstrating that the effective noise level for a DD-protected gate is indeed lower than for an unprotected gate. 
Successfully meeting this challenge would mark a substantial stride towards an accelerated realization of FTQC.

Parallel to the progress in hardware and algorithms, software libraries \cite{StrawberryFields1,StrawberryFields2,Stavenger2022Bosonic-Qiskit,PennyLane} have been developed that can numerically emulate bosonic quantum devices on classical computers. 
However, these software capabilities are still at a nascent stage compared to their qubit counterparts, and still have room for growth. 
Publicizing software libraries alongside comprehensive documentation with tutorials will contribute to the expansion of the bosonic quantum computing community, expediting the availability and usefulness of this quantum platform for applications within and beyond chemistry.

In this perspective, we have showcased recent work that demonstrates the potential for simulating chemistry on bosonic quantum devices and have outlined what we believe would be fruitful near-term applications of bosonic devices to the chemical sciences. 
Pursuing such applications will expand the range of applications of bosonic quantum devices and establish qumode-based platforms as viable and attractive approaches for studying chemistry. Ultimately, the goal of such an approach is to demonstrate a quantum speedup advantage over the best available, or even the best possible, classical methods. 
Given the advances described in this work, we believe that bosonic devices represent an attractive prospect for a demonstration of this type in the near term.


\begin{acknowledgement}

We acknowledge support from the NSF-CCI  Center for Quantum Dynamics on Modular Quantum Devices (CQD-MQD) under grant number CHE-2124511.
Heidi Hendrickson acknowledges computational resources provided in part by the MERCURY consortium under NSF grants CHE-1229354, CHE-1662030, and CHE-2018427.

\end{acknowledgement}


\bibliography{CCI}


\begin{tocentry}

\begin{center}

\includegraphics[width=0.9\columnwidth]{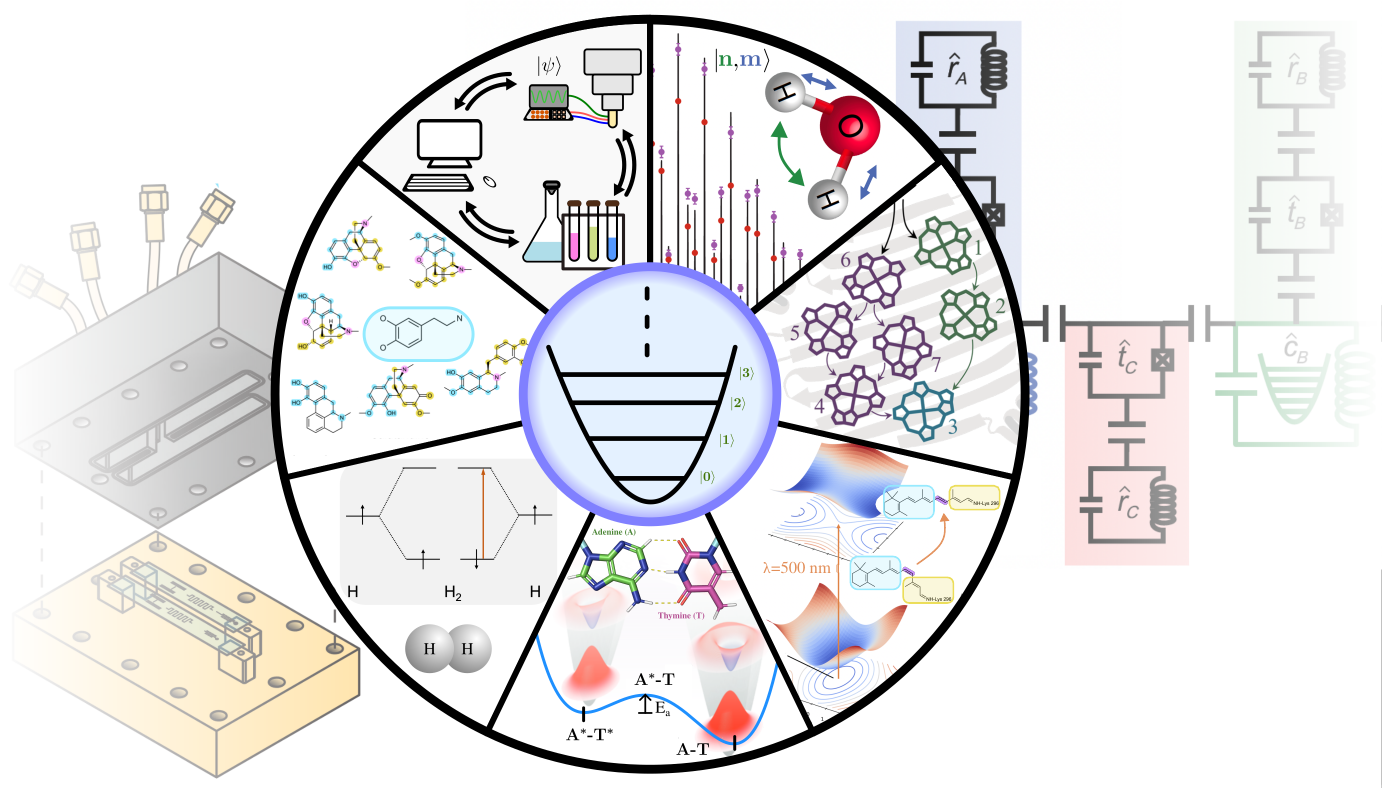}
    
\end{center}

\end{tocentry}


\end{document}